%% file: DEUXPHASES.tex
\documentclass[eqno, 12pt]{amsart}
\usepackage{amssymb, amsmath,  amscd, graphicx, color}

\newtheorem{lemma}{Lemma}[section]
\newtheorem{proposition}{Proposition}[section]
\newtheorem{theorem}{Theorem}[section]

\newtheorem{definition}{Definition}[section]
\newtheorem{remark}{Remark}[section]

\def\ra{\rightarrow}

\newcommand{\C}{{\mathbb C}}

\newcommand{\N}{{\mathbb N}}

\newcommand{\R}{{\mathbb R}}

\newcommand{\U}{{\mathbb U}}

\newcommand{\Z}{{\mathbb Z}}

\def\cC{{\mathcal C}}

\def\cH{{\mathcal H}}

\def\cU{{\mathcal U}}

\newcommand{\diff}{\mathrm{d}}
\def\tg{{\tt g}}
\def\tG{{\tt G}}
\newcommand{\ga}{\Gamma}
\newcommand{\gq}{\Gamma^q}
\newcommand{\gd}{\Gamma^2}
\newcommand{\uq}{u_\Lambda(\Gamma^q)}
\newcommand{\ud}{u_\Lambda(\Gamma^2)}
\newcommand{\ig}{\mathrm{Int}\,\Gamma}
\newcommand{\igm}{\mathrm{Int}_m\,\Gamma}

\newcommand{\igmq}{\mathrm{Int}_m\,\Gamma^q}
\newcommand{\igud}{\mathrm{Int}_1\,\Gamma^2}

\newcommand{\vgm}{V_m(\Gamma)}

\newcommand{\vq}{V(\Gamma^q)}
\newcommand{\dv}{\partial V}

\newcommand{\eg}{\mathrm{Ext\,}\Gamma}
\newcommand{\sg}{\mathrm{supp\,}\Gamma}
\newcommand{\sgq}{\mathrm{supp\,}\Gamma^q}
\newcommand{\sgd}{\mathrm{supp\,}\Gamma^2}

\newcommand{\vfi}{\varphi}

\newcommand{\tq}{\Theta_q}

\newcommand{\tim}{\Theta_m(\mathrm{Int}_m\,\Gamma^q)}
\newcommand{\tiq}{\Theta_q(\mathrm{Int}_m\,\Gamma^q)}

\begin{document}

\vspace*{1cm}
\centerline{\Large\bf On the Singularity of the Free Energy at First}
\vskip 5pt
\centerline{\Large\bf  Order Phase Transition}
\vspace*{1cm}
\centerline{S. Friedli}
\centerline{Institut de physique th\'eorique}
\centerline{EPF-L CH-1015 Lausanne Switzerland}
\bigskip
\centerline{Ch.-Ed. Pfister}
\centerline{Institut de math\'ematiques}
\centerline{EPF-L CH-1015 Lausanne Switzerland}
\bigskip
\vspace*{1cm}

\textbf{Abstract:}
At first order phase transition the
free energy does not  have an analytic continuation in the 
thermodynamical variable, which is conjugate to an order parameter 
for the transition. This result is proved at low temperature for 
lattice models with finite range interaction and two periodic 
ground-states, under the only condition that they verify Peierls condition.
\section{Introduction}\label{section1}
\setcounter{equation}{0}

We study a  lattice model with finite state space on $\Z^d$, $d\geq 2$.  Let 
$\cH_0$ be a
Hamiltonian with finite-range periodic interaction, having  two 
periodic ground-states $\psi_1$ and $\psi_2$,  and so that Peierls condition
is verified.  Let  $\cH_1$ be a  Hamiltonian with periodic and finite range  
interaction, so that the perturbed Hamiltonian
$$
\cH^\mu=\cH_0+\mu\cH_1
$$
splits the degeneracy of the ground-states of $\cH_0$: if  $\mu<0$, 
then $\cH^\mu$ has a unique ground-state $\psi_2$, and if $\mu>0$, then
$\cH^\mu$ has a unique ground-state $\psi_1$. The free energy of the model
with Hamiltonian $\cH^\mu$, at inverse temperature $\beta$, is denoted 
by $f(\mu,\beta)$. Our main result is

\begin{theorem}\label{thm1.1}
Under the above setting, there exist an open interval $U_0\ni 0$, 
$\beta^*\in\R^+$ and, for all $\beta\geq \beta^*$,
$\mu^*(\beta)\in U_0$ with the
following properties.
\begin{enumerate}
\item There is a first-order phase transition  at
$\mu^*(\beta)$. 

\item The free energy $f(\mu,\beta)$ is
analytic in $\mu$ in $\{\mu\in U_0:\,\mu<\mu^*(\beta)\}$; it has a
$C^\infty$ continuation in $\{\mu\in U_0:\,\mu\leq\mu^*(\beta)\}$.

\item The free energy $f(\mu,\beta)$ is analytic in $\mu$ in 
$\{\mu\in U_0:\,\mu>\mu^*(\beta)\}$; it has a $C^\infty$ continuation 
in $\{\mu\in U_0:\,\mu\geq\mu^*(\beta)\}$.

\item There is no analytic continuation of  $f$ from 
$\mu<\mu^*(\beta)$ to $\mu>\mu^*(\beta)$ across $\mu^*(\beta)$, 
or vice-versa.
\end{enumerate}
\end{theorem}

This theorem answers a fundamental theoretical question:
does the free energy, which is analytic in the region of  
a single phase, have an analytic 
continuation beyond a first-order phase transition point? 
The answer is yes for the theory of a simple fluid of van 
der Waals or for mean-field theories. 
The analytic continuation of the free energy beyond the transition 
point was interpreted as the free energy of a metastable phase. 
The answer is no for models with finite range interaction,
under very general conditions, as Theorem \ref{thm1.1} 
shows. This contrasted behavior has its origin in 
the fact that for models with finite range interaction there is 
spatial phase separation at first order phase transition,
contrary to what happens in a mean-field model. Theorem \ref{thm1.1} and its proof
confirm the prediction of the droplet model
\cite{F}. 

Theorem \ref{thm1.1} generalizes the works of Isakov \cite{I1}
for the Ising model and \cite{I2}, where a similar theorem
is proven under  additional assumptions, which are not easy to
verify in a concrete model.  
Our version of Theorem \ref{thm1.1}, which relies uniquely on Peierls condition, is therefore a genuine improvement of \cite{I2}. 
The first result of this kind was proven by Kunz and Souillard \cite{KS}; it concerns the non-analytic behavior of the generating function of the cluster size distribution in percolation, which plays the role of a free energy in that model.
The first statement of Theorem \ref{thm1.1} is a particular case of the theory of Pirogov and Sinai
(see  \cite{S}). We give a  proof of this result, as far as it concerns the free energy,
since we need detailed informations about the phase diagram in the complex plane of the parameter $\mu$.  

The obstruction to an analytic
continuation of the free energy in the variable $\mu$ is due to 
the stability of the droplets of both phases in a neighborhood of $\mu^*$. 
Our proof follows for the essential that of
Isakov in \cite{I1}. We give
a detailed proof of Theorem \ref{thm1.1}, and do not
assume any familiarity with \cite{I1} or \cite{I2}. On the other
hand we assume that the reader is familiar with the cluster
expansion technique. 

The results presented here  are true for a much larger class of systems, but for
the sake of simplicity we restrict our discussion in that paper to the above 
setting, which is already quite general. For example, Theorem \ref{thm1.1}
is true for Potts model with high number $q$ of components at the first order phase transition point $\beta_c$, where the $q$ ordered phases coexist with the disordered phase. Here $\mu=\beta$, the inverse temperature, and the statement is that the free energy, which is analytic for $\beta>\beta_c$, or for $\beta<\beta_c$, does not have an analytic continuation across $\beta_c$. Theorem \ref{thm1.1} is also true when the model has more than two ground-states. For example, for the Blume-Capel model, whose Hamiltonian is
$$
-\sum_{i,j}(s_i-s_j)^2-h\sum_is_i-\lambda\sum_i s_i^2\quad 
\text{with}\quad s_i\in\{-1,0,1\}\,,
$$
the free energy is an analytic function of $h$ and $\lambda$ in the single phase regions. At low temperature, at the triple point occurring at $h=0$ and $\lambda=\lambda^*(\beta)$ there is no analytic continuation of the free energy in $\lambda$, along the path $h=0$, or in the variable  $h$, along the path $\lambda=\lambda^*$. The case of coexistence of more than two phases will be treated in a separate paper.

In the rest of the section we fix the main notations following chapter 
two of Sinai's book \cite{S}, so that the reader may easily find more
information if necessary. We also state Lemma \ref{lem1.1} which contains all estimates on partition functions or free energies. We omit the proof, which 
relies on the cluster expansion method. 

The model is defined on the
lattice $\Z^d$, $d\geq 2$. The spin variables $\vfi(x)$,
$x\in\Z^d$, take values in a finite state space.  If $\vfi,\psi$
are two spin configurations, then $\vfi=\psi$ (a.s.) means that
$\vfi(x)\not =\psi(x)$ holds only on a finite subset of $\Z^d$.
The restriction of $\vfi$ to a subset $A\subset\Z^d$ is denoted by
$\vfi(A)$. The cardinality of a subset $S$ is denoted by $|S|$. If
$x,y\in\Z^d$, then $|x-y|:=\max_{i=1}^d|x_i-y_i|$;  if $W\subset
\Z^d$ and $x\in\Z^d$, then $d(x,W):=\min_{y\in W}|x-y|$ and
if $W, W^\prime$ are subsets of $\Z^d$, then 
$d(W,W^\prime)=\min_{x\in W}d(x,W^\prime)$. We define for $W\subset\Z^d$
$$
\partial W:=\{x\in W:\,d(x,\Z^d\backslash W)=1\}\,.
$$ 
A subset $W\subset\Z^d$ is connected if any two points $x,y\in W$
are connected by a path $\{x_0,x_1,\ldots,x_n\}\subset W$, with
$x_0=x$, $x_n=y$ and $|x_i-x_{i+1}|=1$, $i=0,1,\ldots,n-1$. A
component is a maximally connected subset.

Let $\cH$ be a Hamiltonian with finite-range and periodic bounded
interaction. By introducing an equivalent model on a sublattice,
with a larger state space, we can assume that the  model is
translation invariant with  interaction between neighboring spins
$\varphi(x)$ and $\varphi(y)$, $|x-y|=1$, only.
Therefore, without restricting the generality,  we assume that this is the case
and that the interaction is $\Z^d$-invariant.  
The Hamiltonian is written
$$
\cH^\mu=\cH_0+\mu\cH_1\,,\quad \mu\in\R\,.
$$
$\cH_0$ has two $\Z^d$-invariant ground-states $\psi_1$
and $\psi_2$, and the perturbation $\cH_1$ splits the
degeneracy of the ground-states of $\cH_0$. We assume that the energy (per unit spin) of the ground-states of $\cH_0$ is $0$. $\cU^\mu_x(\vfi)\equiv\cU_{0,x}+\mu\,\cU_{1,x}$ is the interaction 
energy of the spin located
at $x$ for the configuration $\vfi$, so that by definition
$$
\cH^\mu(\vfi)=\sum_{x\in\Z^d}\cU^\mu_x(\vfi)\quad\text{(formal sum)}\,.
$$
$\cU_{1,x}$ is an order parameter for the phase transition.
If $\vfi$ and $\psi$ are two configurations and $\vfi=\psi$ (a.s.),
then
$$
\cH^\mu(\vfi|\psi):=
\sum_{x\in\Z^d}\big(\cU^\mu_x(\vfi)-\cU^\mu_x(\psi)\big)\,.
$$
This last sum is finite since only finitely many  terms are
non-zero. The main condition, which we impose on $\cH_0$, is  
Peierls condition for the ground-states $\psi_1$ and $\psi_2$. 
Let $x\in\Z^d$ and
$$
W_1(x):=\{y\in\Z^d:\,|y-x|\leq 1\}\,.
$$
The boundary $\partial \vfi$ of the configuration $\vfi$  is the 
subset of $\Z^d$ defined by
$$
\partial\vfi:=\bigcup_{x\in\Z^d}\big\{W_1(x):\,\vfi(W_1(x))\not=
\psi_m(W_1(x))\,,\;m=1,2\big\}\,.
$$
Peierls condition means that there exists a positive constant $\rho$ 
such that for $m=1,2$
$$
\cH_0(\vfi|\psi_m)\geq \rho|\partial \vfi|\quad 
\text{$\forall$ $\vfi$ such that $\vfi=\psi_m$ (a.s.)}\,. 
$$
We shall not write usually the $\mu$-dependence of
some quantity; we write for example $\cH$ or $\cU_x$ instead of
$\cH^\mu$ or $\cU^\mu_x$. 

\begin{definition}
Let $M$ denote a finite connected subset of $\Z^d$, and let $\vfi$
be a configuration. Then a couple $\ga=(M,\vfi(M))$ is called a
contour of $\vfi$ if $M$ is a component of the boundary
$\partial\vfi$ of $\vfi$. A couple $\ga=(M,\vfi(M))$ of this type
is called a contour if there exists at least one configuration
$\vfi$ such that $\ga$ is a contour of $\vfi$.
\end{definition}

If $\ga=(M,\vfi(M))$ is a contour, then $M$ is the support of
$\ga$, which we also denote by $\sg$. Suppose that
$\ga=(M,\vfi(M))$ is a contour and consider the components
$A_\alpha$ of $\Z^d\backslash M$.  Then for each component
$A_\alpha$ there exists a unique ground-state $\psi_{q(\alpha)}$,
such that for each $x\in\partial A_\alpha$ one has
$\vfi(W_1(x))=\psi_{q(\alpha)}(W_1(x))$. The index $q(\alpha)$ is
the label of the component $A_\alpha$. For any contour $\ga$ there
exists a unique infinite component of $\Z^d\backslash\sg$, $\eg$,
called the exterior of $\ga$; all other components are called
internal components of $\ga$. The ground-state corresponding to
the label of $\eg$ is the boundary condition of $\ga$; the
superscript $q$ in $\gq$ indicates that $\ga$ is a contour with
boundary condition $\psi_q$. $\igm$ is the union of all internal
components of $\ga$ with label $m$; $\ig:=\bigcup_{m=1,2}\igm$ is
the interior of $\ga$. We use the abbreviations $|\ga|:=|\sg|$ and
$\vgm:=|\igm|$. We define\footnote{Here our convention differs
from \cite{S}.}
\begin{equation}\label{volume}
V(\gq):=V_m(\gq)\quad m\not =q\,.
\end{equation}
For $x\in\Z^d$, let 
$$
c(x):=\big\{y\in\R^d:\,\max_{i=1}^d|x_i-y_i|\leq 1/2\big\}
$$
be the unit cube of center $x$ in $\R^d$. If $\Lambda\subset\Z^d$,
then $|\Lambda|$ is equal to the 
$d$-volume of 
\begin{equation}\label{2.3}
\bigcup_{x\in\Lambda}c(x)\subset\R^d\,.
\end{equation}
The $(d-1)$-volume of the boundary of the set \eqref{2.3} is denoted by 
$\partial|\Lambda|$. We have 
\begin{equation}\label{2.4}
2d\,|\Lambda|^{\frac{d-1}{d}}\leq \partial|\Lambda|\,.
\end{equation}
The equality in \eqref{2.4} is true for cubes only. When 
$\Lambda=\igm^q$, $m\not=q$, $\vq\equiv|\Lambda|$ and
$\dv(\gq)\equiv\partial|\Lambda|$; there 
exists a positive constant $C_0$ such that
\begin{equation}\label{2.5}
\dv(\gq)\leq C_0|\gq|\quad q=1,2\,.
\end{equation}
For each contour $\ga=(M,\vfi(M))$ there corresponds a unique
configuration $\vfi_{\ga}$ with the properties:
$\vfi_{\ga}=\psi_q$ on $\eg$, where $q$ is the label of $\eg$,
$\vfi_{\ga}(M)=\vfi(M)$, $\vfi_{\ga}=\psi_m$ on $\igm$, $m=1,2$.
$\ga$ is the only contour of $\vfi_{\ga}$. Let $\Lambda\subset
\Z^d$; the notation  $\ga\subset\Lambda$ means that
$\sg\subset\Lambda$, $\ig\subset\Lambda$ and $d(\sg,\Lambda^c)>1$.
A contour $\ga$ of a configuration $\vfi$ is an external contour
of $\vfi$ if and only if $\ga\subset \eg^\prime$ for any contour
$\ga^\prime$ of $\vfi$.

\begin{definition}
Let $\Omega(\gq)$ be the set of configurations $\vfi=\psi_q$
(a.s.) such that $\gq$ is the only external contour of $\vfi$.
Then
$$
\Theta(\gq):=\sum_{\vfi\in\Omega(\gq)}\exp\big[-\beta\cH(\vfi|\psi_q)\big]\,.
$$
Let $\Lambda\subset\Z^d$ be a finite subset; let
$\Omega_q(\Lambda)$ be the set of configurations $\vfi=\psi_q$
(a.s.) such that $\ga\subset\Lambda$ whenever $\ga$ is a contour
of $\vfi$. Then
$$
\Theta_q(\Lambda):=
\sum_{\vfi\in\Omega_q(\Lambda)}\exp\big[-\beta\cH(\vfi|\psi_q)\big]\,.
$$
\end{definition}

Two fundamental identities relate the partition functions
$\Theta(\gq)$ and $\Theta_q(\Lambda)$.
\begin{equation}\label{1.1}
\Theta_q(\Lambda)=\sum\prod_{i=1}^n\Theta(\gq_i)\,,
\end{equation}
where the sum is over the set of all families
$\{\gq_1,\ldots,\gq_n\}$ of external contours in $\Lambda$, and
\begin{equation}\label{1.2}
\Theta(\gq)=
\exp\big[-\beta\cH(\vfi_{\gq}|\psi_q)\big]\prod_{m=1}^2\Theta_m(\igmq)\,.
\end{equation}
We define (limit in the sense of van Hove)
$$
g_q:=\lim_{\Lambda\uparrow\Z^d}-\frac{1}{\beta|\Lambda|}\log\Theta_q(\Lambda)\,.
$$
The energy (per unit volume) of  $\psi_m$ for the Hamiltonian $\cH_1$ is 
$$
h(\psi_m):=\cU_{1,x}(\psi_m)\,.
$$
By definition of $\cH_1$, $h(\psi_2)-h(\psi_1)\not=0$, and we assume that
$$
\Delta:=h(\psi_2)-h(\psi_1)>0\,.
$$
The free energy in the thermodynamical  limit is 
\begin{equation}\label{1.4}
f=
\lim_{\Lambda\uparrow\Z^d}-\frac{1}{\beta|\Lambda|}\log\Theta_q(\Lambda)
+\lim_{\Lambda\uparrow\Z^d}\frac{1}{|\Lambda|}\sum_{x\in\Lambda}\cU_x(\psi_q)
=g_q +\mu\,h(\psi_q) \,.
\end{equation}
It is independent of the boundary condition $\psi_q$.

\begin{definition}
Let $\gq$ be a contour with boundary condition $\psi_q$. The
weight $\omega(\gq)$ of $\gq$ is
$$
\omega(\gq):=\exp\big[-\beta\cH(\vfi_{\gq}|\psi_q)\big]\prod_{m:m\not
=q} \frac{\Theta_m(\igmq)}{\Theta_q(\igmq)}\,.
$$
The (bare) surface energy of a contour $\gq$ is
$$
\|\gq\|:=\cH_0(\vfi_{\ga^q}|\psi_q)\,.
$$ 
\end{definition}

For a contour $\gq$ we set
$$
a(\vfi_{\gq}):=\sum_{x\in\sgq}\cU_{1,x}(\vfi_{\gq})-\cU_{1,x}(\psi_q)\,.
$$
Since the interaction is bounded, there exists a constant $C_1$ so that
\begin{equation}\label{2.1}
|a(\vfi_{\gq})|\leq C_1|\gq|\,.
\end{equation}
Using these notations we have
\begin{align}\label{2.2}
\cH(\vfi_{\gq}|\psi_q)
&=
\sum_{x\in\sgq}\big(\cU_x(\vfi_{\gq})-\cU_x(\psi_q)\big)+
\sum_{x\in{\ig}^q}\big(\cU_x(\vfi_{\gq})-\cU_x(\psi_q)\big)\nonumber\\
&=
\cH_0(\vfi_{\gq}|\psi_q)+\mu a(\vfi_{\gq})
+\mu(h(\psi_m)-h(\psi_q))\vq \nonumber\\
&=
\|\gq\|+\mu a(\vfi_{\gq})+\mu (h(\psi_m)-h(\psi_q)) \vq\quad (m\not=q)\,.
\end{align}
The surface energy $\|\ga^q\|$ is always strictly positive since Peierls condition holds, and there exists a constant $C_2$, independent of 
$q=1,2$, such that
\begin{equation}\label{C2}
\rho |\ga^q|\leq\|\ga^q\|\leq C_2|\ga^q|\,.
\end{equation}

\begin{definition}
The weight $\omega(\gq)$ is $\tau$-stable for $\gq$ if
$$
|\omega(\gq)|\leq\exp(-\tau|\gq|)\,.
$$
\end{definition}

For finite subset $\Lambda\subset\Z^d$, using \eqref{1.1} and
\eqref{1.2}, one obtains easily the following identity for the
partition function $\Theta_q(\Lambda)$,
\begin{equation}\label{1.3}
\Theta_q(\Lambda)=1+\sum\prod_{i=1}^n \omega(\gq_i)\,,
\end{equation}
where the sum is over all families of compatible contours
$\{\gq_1,\ldots,\gq_n\}$ with boundary condition $\psi_q$, that is,
$\gq_i\subset\Lambda$ and $d(\sgq_i,\sgq_j)>1$ for all $i\not= j$,
$i,j=1,\ldots,n$, $n\geq 1$. If the weights of all contours with 
boundary condition $\psi_q$ are $\tau$-stable and if $\tau$ is large
enough, then one can express the logarithm of $\Theta_q(\Lambda)$
as an absolutely convergent sum, 
\begin{equation}\label{clusterexpansion}
\log\Theta_q(\Lambda)= \sum_{m\geq
1}\frac{1}{m!}\sum_{\gq_1\subset\Lambda}\cdots\sum_{\gq_m\subset\Lambda}
\varphi_m^T(\gq_1,\ldots,\gq_m)\prod_{i=1}^m\omega(\gq_i)\,.
\end{equation}
In \eqref{clusterexpansion} $\varphi_m^T(\gq_1,\ldots,\gq_m)$ is a
purely combinatorial factor. This is the basic formula which is used for 
controlling $\Theta_q(\Lambda)$. We also introduce restricted partition
functions and free energies. 
For each $n=0,1,\ldots,$ we define new weights $\omega_n(\gq)$
$$
\omega_{n}(\gq):=
\begin{cases}
\omega(\gq) & \text{if 
$\vq\leq n$,}\\
0 & \text{otherwise.}
\end{cases}
$$
For $q=1,2$, we define $\tq^n$ by equation \eqref{1.3}, replacing $\omega(\gq)$ by $\omega_n(\gq)$, and we set (provided that 
$\Theta^{n}_q(\Lambda)\not=0$ for all $\Lambda$)
\begin{equation}\label{stablefree}
g_q^{n}:=
-\lim_{\Lambda\uparrow\Z^d}\frac{1}{\beta|\Lambda|}
\log\Theta^{n}_q(\Lambda)\quad\text{and}
\quad
f_q^n:=g_q^n+z\,h(\psi_q)\,.
\end{equation}
$f_q^n$ is the restricted free energy of order $n$ and boundary condition 
$\psi_q$. Let
\begin{equation}\label{l0}
l(n):=C_0^{-1}\big\lceil 2d n^{\frac{d-1}{d}}\big \rceil \quad n\geq 1\,.
\end{equation}
Notice that $\Theta^{n}_q(\Lambda)=\Theta_q(\Lambda)$ if 
$|\Lambda|\leq n$, and that $V(\gq)\geq n$ implies that 
$|\gq|\geq l(n)$ since \eqref{2.4} and \eqref{2.5} hold.

\begin{lemma}\label{lem1.1}
Suppose that the weights $\omega(\gq)$ are $\tau$-stable for all $\gq$. 
Then there exists $K_0<\infty$ and $\tau^*_0<\infty$, so that
for all $\tau\geq\tau^*_0$, \eqref{clusterexpansion} is absolutely 
convergent and
$$
\beta|g_q|\leq K_0{\rm e}^{-\tau}\,.
$$
For all subsets $\Lambda\subset\Z^d$,
$$
\big|\log \tq(\Lambda)+\beta g_q\,|\Lambda|\big|\leq 
K_0{\rm e}^{-\tau}\,\partial|\Lambda|\,.
$$
If  $\omega(\gq)=0$ for all $\gq$ such that $|\gq|\leq m$, then
$$
\beta |g_q|\leq \big(K_0{\rm e}^{-\tau}\big)^m\,.
$$
For $n\geq 1$ and $m\geq n$,
$$
\beta |g_q^m-g_q^{n-1}|\leq \big(K_0{\rm e}^{-\tau}\big)^{l(n)}\,.
$$
Furthermore, if $\omega(\gq)$ depends on a parameter $t$ and 
$$
\big|\frac{d}{dt}\omega(\gq)\big|\leq  D_1{\rm e}^{-\tau|\gq|}\quad\text{and}\quad
\big|\frac{d^2}{dt^2}\omega(\gq)\big|\leq  D_2{\rm e}^{-\tau|\gq|}\,,
$$
then there exists $K_k<\infty$ and $\tau^*_k<\infty$, $k=1,2$, 
so that for all $\tau\geq\tau^*_k$, 
$\displaystyle\frac{d^k}{dt^k}g_q$ exists and 
$$
\beta\big|\frac{d}{dt}g_q\big|\leq D_1 K_1{\rm e}^{-\tau}
\quad\text{and}\quad
\beta\big|\frac{d^2}{dt^2}g_q\big|\leq \max\{D_2,D_1^2\}
K_2{\rm e}^{-\tau}\,.
$$
For all subsets $\Lambda\subset\Z^d$,
$$
\big|\frac{d}{dt}\log \tq(\Lambda)
+\beta \frac{d}{dt}g_q\,|\Lambda|\big|\leq 
D_1K_1{\rm e}^{-\tau}\,\partial|\Lambda|
$$
and
$$
\big|\frac{d^2}{dt^2}\log \tq(\Lambda)
+\beta \frac{d^2}{dt^2}g_q\,|\Lambda|\big|\leq 
\max\{D_2,D_1^2\}K_2{\rm e}^{-\tau}\,\partial|\Lambda|\,.
$$
\end{lemma}

Lemma \ref{lem1.1} is proved by the cluster expansion method.
It follows from \eqref{clusterexpansion} and arguments similar to those
of the proof of Lemma 3.5. in
section 3.3 in \cite{Pf}.

\section{Proof of Theorem \ref{thm1.1}}\label{section2}
\setcounter{equation}{0}

The  proof of Theorem \ref{thm1.1} is given in the next five
subsections. In subsection \ref{subsection2.1} we construct the phase
diagram and in subsection \ref{subsection2.2} we study the analytic continuation of the weights of contours
in a neighborhood of the point of phase coexistence $\mu^*$. 
The results about the analytic continuation are crucial for the rest of 
the analysis. Construction of the phase diagram in the complex plane has been done by Isakov \cite{I2}.
We follow partly this reference and Zahradnik \cite{Z}.
In subsection \ref{subsection2.3} we derive an expression of the derivatives of the free energy at finite volume. We prove a lower bound for a restricted class
of terms of this expression. This is an improved version  of a
similar analysis of Isakov \cite{I1}. From these results we obtain a lower bound for the derivatives  of 
the free energy $f_\Lambda$ in a finite box $\Lambda$. We show in
subsection \ref{subsection2.4} that for
large $\beta$, there exists an increasing diverging sequence
$\{k_n\}$, so that the $k_n^{th}$-derivative of $f_\Lambda$ with respect
to $\mu$, evaluated at $\mu^*$, behaves as $k_n!^{\frac{d}{d-1}}$
(provided that $\Lambda$ is large enough). In the last subsection we 
end the proof of the  impossibility of an
analytic continuation of the free energy across $\mu^*$, by showing that the results of subsection \ref{subsection2.4} remain true in the thermodynamical limit.

\subsection{Construction of the phase diagram in the complex plane}
\label{subsection2.1}

We construct the phase diagram  for complex values of the parameter $\mu$, by
constructing iteratively the phase diagram for the restricted free energies
$f^n_q$ (see \eqref{stablefree}).
We set $z:=\mu+i\nu$. The method consists in finding a sequence of intervals
for each $\nu\in\R$,
$$
U_n(\nu;\beta):=(\mu^*_n(\nu;\beta)-b^1_n,\mu^*_n(\nu;\beta)+b^2_n)\,,
$$ 
with the properties
\begin{equation}\label{inclusion}
(\mu^*_n(\nu;\beta)-b^1_n,\mu^*_n(\nu;\beta)+b^2_n)\subset
(\mu^*_{n-1}(\nu;\beta)-b^1_{n-1},\mu^*_{n-1}(\nu;\beta)+b^2_{n-1})
\end{equation}
and $\lim_nb^q_n=0$, $q=1,2$. By construction of the intervals
$U_{n-1}(\nu;\beta)$ the restricted free energies $f_{q}^{n-1}$ of order $n-1$,
$q=1,2$, are well-defined and analytic on 
$$
\U_{n-1}:=\{z\in\C:\,\mathrm{Re}z\in U_{n-1}(\mathrm{Im}z;\beta)\}\,.
$$ 
The point $\mu_n^*(\nu;\beta)$, $n\geq 1$, is solution of the equation 
$$
\mathrm{Re}\big(f_2^{n-1}(\mu_{n}^*(\nu;\beta)+i\nu)-
f_1^{n-1}(\mu_{n}^*(\nu;\beta)+i\nu)\big)=0\,.
$$
$\mu_n^*(0;\beta)$ is the point of phase coexistence for the restricted free energies of order $n-1$, and the point of phase coexistence of the model is given by $\mu^*(0;\beta)=\lim_n\mu^*_n(0;\beta)$. 
This iterative procedure also gives the necessary results needed in subsection \ref{subsection2.1} about the analytic continuation of the weights $\omega(\gq)$  around the point of phase coexistence $\mu^*$. Since we need sharp results about the analytic continuation of the weights $\omega(\gq)$, we must choose carefully the two sequences $\{b^q_n\}$,
$q=1,2$. In order to ease the exposition we first describe the iterative 
procedure with a specific choice of $\{b^q_n\}$,  based on the 
isoperimetric inequality
\begin{equation}\label{iso12}
V(\gq)^{\frac{d-1}{d}}\leq\chi^{-1}\|\gq\|\quad\forall\;\gq\,,\,q=1,2\,.
\end{equation}
Existence of $\chi$ in \eqref{iso12}  follows from \eqref{2.4}, \eqref{2.5} and \eqref{C2}. 
Then, in subsection \ref{subsection2.2}, we make another choice for $\{b^q_n\}$.
This iterative construction is given in details in the proof of the Proposition \ref{pro2.1}, which is the main result of subsection \ref{subsection2.1}.

\begin{proposition}\label{pro2.1}
Let $0<\varepsilon<\rho$ and $0<\delta<1$ so that $\Delta-2\delta>0$.
Set
$$
U_0:=(-C_1^{-1}\varepsilon,C_1^{-1}\varepsilon)
\quad\text{and}\quad
\U_0:=\{z\in\C:\,\mathrm{Re}z\in U_0\}\,
$$ 
and
$$
\tau(\beta):=\beta(\rho-\varepsilon)-3C_0\delta\,.
$$
There exists $\beta_0\in\R^+$ such that for all $\beta\geq\beta_0$ the following holds. \\
A) There exists a continuous real-valued function on $\R$, 
$\nu\mapsto\mu^*(\nu;\beta)$, so that
$\mu^*(\nu;\beta)+i\nu\in\U_0$. \\
B) If $\mu+i\nu\in\U_0$ and $\mu\leq \mu^*(\nu;\beta)$, then the weight
$\omega(\ga^2)$ is $\tau(\beta)$-stable for all contours $\ga^2$
with boundary condition $\psi_2$, and
analytic in $z=\mu+i\nu$ if $\mu< \mu^*(\nu;\beta)$.\\
C) If $\mu+i\nu\in\U_0$ and $\mu\geq \mu^*(\nu;\beta)$, then the weight
$\omega(\ga^1)$ is $\tau(\beta)$-stable for all contours $\ga^1$
with boundary condition $\psi_1$, and
analytic in $z=\mu+i\nu$ if $\mu >\mu^*(\nu;\beta)$.
\end{proposition}

\begin{remark}{\rm
$\rho$ is the constant of the Peierls condition and 
$\Delta=h(\psi_2)-h(\psi_1)>0$. 
We may choose $\delta$ in such a way that  $\delta=\delta(\beta)$ and $\lim_{\beta\ra\infty}\delta(\beta)=0$, without changing the theorem. Indeed, the only condition which we need to satisfy is \eqref{delta}. So, whenever we need it, we consider $\delta$ as function of $\beta$, so that by taking $\beta$ large enough, 
we have $\delta$ as small as we wish. }
\end{remark}

\begin{proof}
The iterative method depends on a free parameter $\theta^\prime$,
$0<\theta^\prime<1$.
On the interval $U_0(\nu;\beta):=(-b_0,b_0)$ with $b_0=\varepsilon C_1^{-1}$, 
$f_q^0(\mu+i\nu)$ is defined and we set $\mu^*_0(\nu;\beta):= 0$. The two  decreasing sequences $\{b^q_n\}$, $q=1,2$ and $n\geq 1$, are defined in \eqref{defsequence}. The iterative construction is possible whenever 
the sequences $\{b^q_n\}$, $q=1,2$, verify \eqref{mc1},
\eqref{mc2} and \eqref{mc3}. We prove iteratively the following statements.
\begin{enumerate}
\item[A.]
$f_q^n(\mu+i\nu;\beta)$ is defined for all $\mu\in U_{n-1}(\nu;\beta)$,
and $\nu\mapsto\mu_n^*(\nu;\beta)$ is a continuous solution of the equation 
$$
\mathrm{Re}\big(f_2^{n-1}(\mu_{n}^*(\nu;\beta)+i\nu)-
f_1^{n-1}(\mu_{n}^*(\nu;\beta)+i\nu)\big)=0\,,
$$
so that \eqref{inclusion} holds. 
\item[B.]
On  $\U_n$, $\omega_n(\gq)$ is analytic  for any contour $\gq$, $q=1,2$, and 
$\omega_n(\gq)$ is $\tau_1(\beta)$-stable (see \eqref{tau1}).
\item[C.]
On  $\U_n$, 
$\displaystyle \big|\frac{d}{dz}\omega_n(\gq)\big|
\leq \beta C_3{\rm e}^{-\tau_2(\beta)|\gq|}$
(see \eqref{tau2} and \eqref{C3}).
\item[D.]
For each $n\geq 1$, if $\mu\leq \mu_{n}^*(\nu;\beta)-b^1_n$, then
$\omega(\ga^2)$ is $\tau(\beta)$-stable for any $\ga^2$ with boundary condition $\psi_2$.
Similarly, for each $n\geq 1$, if $\mu\geq \mu_{n}^*(\nu;\beta)+b^2_n$, then
$\omega(\ga^1)$ is $\tau(\beta)$-stable for any $\ga^1$ with boundary condition $\psi_1$.
\end{enumerate}
From these results the proposition follows with 
$$
\mu^*(\nu;\beta)=\lim_{n\ra\infty}\mu^*_n(\nu;\beta)\,.
$$
The analyticity of the weights $\omega(\ga^q)$ is an immediate consequence of their stability since $\tim$ and $\tiq\not=0$ are analytic. 

Let $0<\theta^\prime<1$ be given, as well as $\varepsilon$ and $\delta$ as in the proposition. We introduce all constants used in the proof below.
\begin{equation}\label{tau1}
\tau_1(\beta;\theta^\prime):=
\beta\big(\rho(1-\theta^\prime)-\varepsilon\big)-2\delta C_0\,,
\end{equation}
\begin{equation}\label{tau2}
\tau_2(\beta;\theta^\prime):=
\tau_1(\beta;\theta^\prime)-\frac{d}{d-1}\,,
\end{equation}
and
\begin{equation}\label{C3}
C_3:=C_1+2\delta C_0+(\Delta+2\delta)(\chi^{-1}C_2)^\frac{d}{d-1}\,.
\end{equation}
We assume that $\beta_0$ is large enough so that\footnote{$\tau^*_k$, $k=0,1,2$, are defined in Lemma \ref{lem1.1}. 
Condition  $\tau_2(\beta)>\tau^*_2$ is needed only in Lemma \ref{lem2.1}. We have stated Lemma \ref{lem2.1} separately in order to simplify the proof
of Proposition \ref{pro2.1}.}
 $\tau_2(\beta)>\max\{\tau^*_0,
\tau^*_1,\tau^*_2\}$,
\eqref{numero} holds, 
\begin{equation}\label{delta}
K{\rm e}^{-\tau_1(\beta)}\leq\delta\quad\text{and}\quad
C_3K{\rm e}^{-\tau_2(\beta)}\leq\delta\,,
\end{equation}
where $K=\max\{K_0,K_1\}$, and $K_0$, $K_1$ are the  constants of Lemma \ref{lem1.1}. 
We assume that for $q=1,2$,
\begin{equation}\label{mc1}
b^q_{n}-b^q_{n+1}> \frac{2\delta^{l(n)}}{\beta(\Delta-2\delta)}\;,\quad \forall n\geq 1\,.
\end{equation}
If we define
\begin{equation}\label{defsequence}
b^1_n\equiv b^2_n:=\frac{\chi\theta^\prime}{(\Delta+2\delta)n^{\frac{1}{d}}}\,,\;
n\geq 1\,,
\end{equation}
then it is immediate to verify \eqref{mc1} when $\beta$ is large enough or $\delta$ small enough. 
On $\U_0$ all contours $\ga$ 
with empty interior are $\beta(\rho-\varepsilon)$-stable
(see \eqref{2.1}), and 
$$
\Big|\frac{d}{dz}\omega(\ga)\Big|\leq 
\beta C_1|\ga|{\rm e}^{-\beta(\rho-\varepsilon)|\ga|}
\leq
\beta C_1{\rm e}^{-[\beta(\rho-\varepsilon)-1]|\ga|}
\leq \beta C_3 {\rm e}^{-\tau_2(\beta)|\ga|}\,.
$$

Assume that the construction has been done for all $m\leq n-1$. 
By Lemma
\ref{lem1.1}, if $z\in\U_{n-1}$, then
\begin{equation}\label{ind}
\big|\frac{d}{dz}g_q^{m}\big|\leq C_3K{\rm e}^{-\tau_2(\beta)}\leq\delta
\quad m\leq n-1\,.
\end{equation}

\bigskip
\noindent
A. We prove the existence of $\mu_n^*(\nu;\beta)\in\U_{n-1}$.
$\mu^*_{n}(\nu;\beta)$ is solution of the equation
$$
\mathrm{Re}\big(f_2^{n-1}(\mu_{n}^*(\nu;\beta)+i\nu)-
f_1^{n-1}(\mu_{n}^*(\nu;\beta)+i\nu)\big)=0\,.
$$
Let $F^{m}(z):=f_2^{m}(z)-f_1^{m}(z)$. Then, for $\mu^\prime+i\nu\in\U_{n-1}$,
\begin{align}\label{222}
F^{n-1}(\mu^\prime+i\nu)
&=
F^{n-1}(\mu^\prime+i\nu)-F^{n-2}(\mu_{n-1}^*+i\nu)\\
&=
F^{n-1}(\mu^\prime+i\nu)-F^{n-1}(\mu_{n-1}^*+i\nu)
+F^{n-1}(\mu_{n-1}^*+i\nu)\nonumber\\
&\quad-F^{n-2}(\mu_{n-1}^*+i\nu)\nonumber\\
&=
\int_{\mu_{n-1}^*}^{\mu^\prime}\!\!\frac{d}{d\mu}F^{n-1}(\mu+i\nu)\,d\mu
+\big(g_2^{n-1}-g_2^{n-2}\big)(\mu_{n-1}^*+i\nu)\nonumber\\
&\quad-
\big(g_1^{n-1}-g_1^{n-2}\big)(\mu_{n-1}^*+i\nu)\,.\nonumber
\end{align}
If $V(\ga)= n-1$, then $|\ga|\geq l(n-1)$.
Therefore, by Lemma \ref{lem1.1},
\begin{equation}\label{estimate}
|\big(g_q^{n-1}-g_q^{n-2}\big)(\mu_{n-1}^*+i\nu)|
\leq \beta^{-1}\delta^{l(n-1)}\,.
\end{equation}
If $z^\prime=\mu^\prime+i\nu\in\U_{n-1}$, then \eqref{222},
\eqref{ind} and \eqref{estimate} imply
\begin{align*}
\Delta(\mu^\prime-\mu_{n-1}^*)
+2\delta|\mu^\prime-\mu_{n-1}^*|&+2\beta^{-1}\delta^{l(n-1)}
\geq \mathrm{Re}F^{n-1}(z^\prime)\\
&\geq 
\Delta(\mu^\prime-\mu_{n-1}^*)
-2\delta|\mu^\prime-\mu_{n-1}^*|-2\beta^{-1}\delta^{l(n-1)}\,.
\end{align*}
\eqref{mc1} implies
$$
b^q_{n-1}>b^q_{n-1}-b^q_{n}> \frac{2\delta^{l(n-1)}}{\beta(\Delta-2\delta)}\,,
$$
so that $\mathrm{Re}F^{n-1}(\mu_{n-1}^*-b^1_{n-1}+i\nu)<0$ and
$\mathrm{Re}F^{n-1}(\mu_{n-1}^*+b^2_{n-1}+i\nu)>0$. This proves the 
existence  of $\mu_n^*$ and its  uniqueness, since 
$\mu\mapsto \mathrm{Re}F^{n-1}(\mu+i\nu)$ is strictly increasing.
Moreover, by putting $\mu^\prime=\mu_{n}^*(\nu;\beta)$ in 
\eqref{222}, we get
$$
|\mu_{n}^*(\nu;\beta)-\mu_{n-1}^*(\nu;\beta)|\leq 
\frac{2\delta^{l(n-1)}}{\beta(\Delta-2\delta)}\,.
$$
Therefore $\U_n\subset \U_{n-1}$. The implicit function theorem implies that
$\nu\mapsto \mu_{n}^*(\nu;\beta)$ is continuous (even $C^\infty$).

\bigskip
\noindent
B. We prove that $\omega_n(\gq)$ is $\tau_1$-stable for all contours $\gq$,
$q=1,2$. Let $\gq$ be a contour with 
$\vq=n$. All contours contributing to $\tim$ and $\tiq$ have volumes smaller than $n-1$, so that for these contours $\omega(\ga)=\omega_{n-1}(\ga)$.
If $z\in\U_{n-1}$ (use \eqref{2.1}, \eqref{2.5} and the definition of
$U_0$), then
\begin{align}\label{stability}
|\omega(\gq)| &= \exp\big[-\beta\mathrm{Re}\cH(\vfi_{\gq}|\psi_q)\big]\,
\Big|\prod_{m:m\not=q} \frac{\Theta_m(\igmq)}{\Theta_q(\igmq)}\Big| \\
&\leq
\exp\Big[-\beta\|\gq\|+\big(\beta\varepsilon+2C_0 \delta\big)|\gq|
-\beta\mathrm{Re}\big(f_m^{n-1}-f_q^{n-1}\big)\vq
\Big]  \nonumber\\
&=
\exp\Big[-\beta\|\gq\|+\big(\beta\varepsilon+2C_0 \delta\big)|\gq|
-\beta\mathrm{Re}\big(f_m^{n-1}-f_q^{n-1}\big)\frac{\vq}{\|\gq\|}\|\gq\|\Big]
\,. \nonumber
\end{align}
Let $\mu\in U_{n-1}(\nu;\beta)$. We prove that  $b^q_n$ verify the following conditions, which imply the $\tau_1$-stability.
\begin{align}
-\mathrm{Re}\big(f_1^{n-1}-f_2^{n-1}\big)\frac{V(\ga^2)}{\|\ga^2\|}
&\leq \theta^\prime \quad\text{if $\mu\leq\mu_n^*+b^2_n$ and $V(\ga^2)=n$,}
\label{mc2}\\
-\mathrm{Re}\big(f_2^{n-1}-f_1^{n-1}\big)\frac{V(\ga^1)}{\|\ga^1\|}
&\leq \theta^\prime \quad\text{if $\mu\geq\mu_n^*-b^1_n$
and $V(\ga^1)=n$.}
\label{mc3}
\end{align}
For the present choice of $\{b^q_n\}$, the isoperimetric inequality \eqref{iso12} implies
$$
\frac{\vq}{\|\gq\|}\leq\frac{V(\gq)^\frac{1}{d}}{\chi}\quad\forall\;q=1,2\,,
$$
and therefore
\begin{align*}
\big|\mathrm{Re}\big(f_m^{n-1}-f_q^{n-1}\big)\big|\frac{\vq}{\|\gq\|}
&=
\Big|\mathrm{Re}\int_{\mu^*_n}^\mu\frac{d}{d\mu}
\big(f_m^{n-1}-f_q^{n-1}\big)\,d\mu\Big|\frac{\vq}{\|\gq\|}\\
&\leq
|\mu-\mu_{n}^*|(\Delta+2\delta)
\frac{\vq}{\|\gq\|}\leq\theta^\prime\,.\nonumber
\end{align*}
Conditions \eqref{mc2} and \eqref{mc3} ensure that on $\U_n$
$$
|\omega(\gq)|\leq
\exp\Big[-\beta\big(\rho(1-\theta^\prime)-\varepsilon-2\beta^{-1}C_0 \delta\big)|\gq|\Big]\,.
$$

\bigskip
\noindent
C. We prove that on   $\U_n$  
$$
\big|\frac{d}{dz}\omega_n(\ga)\big|
\leq \beta C_3{\rm e}^{-\tau_2(\beta)|\ga|}\,.
$$
Let $\vq=n$; from \eqref{2.2}
\begin{align*}
\frac{d}{dz}\omega_n(\gq)
&=
\omega_n(\gq)\Big(-\beta a(\vfi_{\gq})-\beta\big(h(\psi_m)-h(\psi_q)\big)\vq \\
&\quad
+\frac{d}{dz}\big(\log\tim-\log\tiq\big)\Big)\,.
\end{align*}
By Lemma \ref{lem1.1} and the isoperimetric inequality \eqref{2.4} we get
\begin{align}\label{derivative}
\big|\frac{d}{dz}\omega_n(\gq)\big|
&\leq 
\beta |\omega_n(\gq)|
\big(|\gq|(C_1+2\delta C_0)+\vq(\Delta+2\delta)\big)\\
&\leq
\beta C_3|\omega_n(\gq)||\gq|^\frac{d}{d-1}\nonumber\\
&\leq
\beta C_3{\rm e}^{-\tau_2(\beta)|\gq|}\nonumber\,.
\end{align}

\bigskip
\noindent
D. We prove
that $\omega(\ga^2)(z)$ is $\tau(\beta)$-stable for
any contour $\ga^2$ with boundary condition $\psi_2$, if $\mu\leq \mu_{n}^*(\nu;\beta)-b^1_n$. Using the induction hypothesis it is sufficient to prove this statement for
$z=\mu+i\nu\in\U_{n-1}$ and $\mu\leq \mu_{n}^*(\nu;\beta)-b^1_n$. 
If $z=\mu+i\nu\in\U_{n-1}$, then all contours with volume 
$V(\ga)\leq n-1$ are $\tau_1(\beta)$-stable, and for
$\mu\leq\mu^*_n$, $\mu\mapsto \mathrm{Re}(f_1^{n-1}-f_2^{n-1})(\mu+i\nu)$
is strictly decreasing. If $\mu\leq \mu_{n}^*(\nu;\beta)-b^1_n$, then (see \eqref{defsequence} and \eqref{mc1})
\begin{align}\label{c}
\beta\mathrm{Re}(f^{n-1}_1-f_2^{n-1})(\mu+i\nu)
&=
-\beta\int^{\mu^*_n}_\mu\frac{d}{d\mu}
\mathrm{Re}(f_1^{n-1}-f_2^{n-1})(\mu+i\nu)\,d\mu \nonumber\\
&\geq
-\beta\int^{\mu^*_n}_{\mu^*_n-b_n^1}\frac{d}{d\mu}
\mathrm{Re}(f_1^{n-1}-f_2^{n-1})(\mu+i\nu)\,d\mu \nonumber\\
&\geq
\beta b_n^1(\Delta-2\delta)\geq 2\delta^{l(n)}\,.
\end{align}
First suppose that $V(\ga^2)\leq n$. From \eqref{c} and 
\eqref{stability} it follows that $\omega(\ga^2)$ is 
$\beta(\rho-\varepsilon-2\beta^{-1}C_0\delta)$-stable, in particular
$\tau(\beta)$-stable. Moreover, if $|\Lambda|\leq n$, then
\begin{equation}\label{ouf}
\Big|\exp\big[-\beta z(h(\psi_1)-h(\psi_2))|\Lambda|\big]
\frac{\Theta_1(\Lambda)}{\Theta_2(\Lambda)}\Big|
\leq {\rm e}^{3 \delta\partial|\Lambda|}\,.
\end{equation}
Indeed,  all contours inside $\Lambda$ are $\tau_1(\beta)$-stable.
By Lemma \ref{lem1.1}
\begin{align*}
\Big|{\rm e}^{-\beta z(h(\psi_1)-h(\psi_2))|\Lambda|}
\frac{\Theta_1(\Lambda)}{\Theta_2(\Lambda)}\Big|
&\leq 
\big|{\rm e}^{-\beta (zh(\psi_1)-zh(\psi_2)+g_1^{n-1}-g_2^{n-1})|\Lambda|}
\big|\,
{\rm e}^{2\delta\partial|\Lambda|}\\
&=
{\rm e}^{-\beta \mathrm{Re}(f_1^{n-1}(z)-f_2^{n-1}(z))|\Lambda|}
{\rm e}^{2\delta\partial|\Lambda|}\\
&\leq
{\rm e}^{2\delta\partial|\Lambda|}\,.
\end{align*}
To prove point D, we prove by induction on $|\Lambda|$ that \eqref{ouf} holds
for any $\Lambda$. Indeed, if \eqref{ouf} is true and if we set $\Lambda:={\rm Int}_1\ga^2$, then it follows easily from the definition of $\omega(\ga^2)$ and from \eqref{2.2} that $\omega(\ga^2)$ is $\tau(\beta)$-stable.

The argument to prove \eqref{ouf} is due to Zahradnik \cite{Z}. The statement is true for $|\Lambda|\leq n$. Suppose that it is true for 
$|\Lambda|\leq m$, $m>n$, and let $|\Lambda|=m+1$. The induction hypothesis
implies that  $\omega(\ga^2)(z)$ is $\tau(\beta)$-stable
if $V(\ga^2)\leq m$. Therefore
\begin{align*}
\Big|{\rm e}^{-\beta z(h(\psi_1)-h(\psi_2))|\Lambda|}
\frac{\Theta_1(\Lambda)}{\Theta_2(\Lambda)}\Big|
&\leq 
\big|{\rm e}^{-\beta (zh(\psi_1)-zh(\psi_2)-g_2^{m})|\Lambda|}
\Theta_1(\Lambda)\big|
{\rm e}^{\delta\partial|\Lambda|}\,.
\end{align*}
From \eqref{1.1}
$$
\Theta_1(\Lambda)=
\sum\prod_{j=1}^r\Theta(\ga^1_j)\,,
$$
where the sum is over all families 
$\{\ga^1_1,\ldots,\ga^1_r\}$ of compatible external contours in
$\Lambda$. 
We say that an external contour $\ga^1_j$ is large if $V(\ga^1_j)\geq n$. Suppose that the contours $\ga^1_1,\ldots\ga^1_p$ are large and all other contours $\ga^1_{p+1},\ldots
\ga^1_r$ not large. We set 
$$
{\rm Ext}_1^p(\Lambda):=\big(\bigcap_{j=1}^p{\rm Ext}\ga^1_j\big)\cap\Lambda\,.
$$
Summing over all contours which are not large, and using \eqref{1.2}, we get
\begin{align*}
\Theta_1(\Lambda)
&=
\sum\Theta^{n-1}_1\big({\rm Ext}_1^p(\Lambda)\big)\prod_{j=1}^p
\exp\big[-\beta\cH(\vfi_{\ga^1_j}|\psi_1)\big]
\Theta_1({\rm Int}_1\ga^1_j)\Theta_2({\rm Int}_2\ga^1_j)\\
&=
\sum\Theta^{n-1}_1\big({\rm Ext}_1^p(\Lambda)\big)\prod_{j=1}^p
{\rm e}^{-\beta\|\ga_j^1\|-\beta za(\vfi_{\ga_j^1})
+\beta z(h(\psi_1)-h(\psi_2))|{\rm Int}_2\ga^1_j|}
\\
&\;\,\quad\cdot\frac{\Theta_1({\rm Int}_1\ga^1_j)}{\Theta_2({\rm Int}_1\ga^1_j)}
\,\Theta_2({\rm Int}_1\ga^1_j)\Theta_2({\rm Int}_2\ga^1_j)
\,;
\end{align*}
the sums are over all families $\{\ga^1_1,\ldots\ga_p^1\}$ of
compatible external large contours in $\Lambda$. 
All contours which are not large are 
$\tau_1(\beta)$-stable, and we use the cluster expansion to control
$\Theta^{n-1}_1\big({\rm Ext}_1^p(\Lambda)\big)$,  
$\Theta_2({\rm Int}_1\ga^1_j)$ and $\Theta_2({\rm Int}_2\ga^1_j)$.
Notice that $\partial|{\rm Ext}_1^p(\Lambda)|\leq
\partial|\Lambda|+\sum_{j=1}^pC_0|\ga^1_j|$. By Lemma \ref{lem1.1} and the induction hypothesis,
\begin{align*}
\Big|{\rm e}^{-\beta z(h(\psi_1)-h(\psi_2))|\Lambda|}
&
\frac{\Theta_1(\Lambda)}{\Theta_2(\Lambda)}\Big|
\leq 
{\rm e}^{2\delta \partial|\Lambda|}\sum
{\rm e}^{-\beta \mathrm{Re}(f_1^{n-1}-f_2^{n-1}-g_2^m+g_2^{n-1})|{\rm Ext}_1^p(\Lambda)|}\\
&
\cdot\prod_{j=1}^p
{\rm e}^{-(\beta\rho -\beta\varepsilon
+6C_0\delta)|\ga^1_j|}
{\rm e}^{-\beta\mathrm{Re}(f_1^{n-1}-f_2^{n-1}-g_2^m+g_2^{n-1})|\ga^1_j|}\,.
\end{align*}
We define
$$
\hat{\tau}(\beta):=\beta(\rho-\varepsilon)-6C_0\delta\,.
$$
From \eqref{c} and Lemma \ref{lem1.1} we have 
$$
\beta (f_1^{n-1}-f_2^{n-1}-g_2^m+g_2^{n-1})\geq \delta^{l(n)}\,.
$$
Hence,
\begin{align*} 
\Big|{\rm e}^{-\beta z(h(\psi_1)-h(\psi_2))|\Lambda|}
\frac{\Theta_1(\Lambda)}{\Theta_2(\Lambda)}\Big|
&\leq 
{\rm e}^{2\delta \partial|\Lambda|}
\sum {\rm e}^{-\delta^{l(n)}|{\rm Ext}_1^p(\Lambda)|}
\prod_{j=1}^p
{\rm e}^{-(\delta^{l(n)}+\hat{\tau}(\beta))|\ga^1_j|}\,.
\end{align*}
We define
$$
\hat{\omega}(\ga):=
\begin{cases}
{\rm e}^{-(\hat{\tau}(\beta)-C_0\delta)|\ga|} & \text{if $|\ga|\geq l(n)$;}\\
0 & \text{otherwise.}
\end{cases}
$$
Let $\hat{\Theta}(\Lambda)$ be defined by  \eqref{1.3}, replacing $\omega(\gq)$ by $\hat{\omega}(\ga)$, and let
$$
\hat{g}:=\lim_{\Lambda \uparrow\Z^d}
-\frac{1}{\beta|\Lambda|}\log\hat{\Theta}(\Lambda)\,.
$$
We assume that $\beta_0$ is large enough so that for all $\beta\geq\beta_0$,
\begin{equation}\label{numero}
K{\rm e}^{-\hat{\tau}(\beta)}\leq\delta\,,
\end{equation}
where $K$ is the constant of Lemma \ref{lem1.1}. Since 
$\beta |\hat{g}|\leq\delta^{l(n)}$, putting into evidence  a factor
${\rm e}^{\beta\hat{g}|\Lambda|}$, we get
\begin{align}\label{da} 
\Big|{\rm e}^{-\beta z(h(\psi_1)-h(\psi_2))|\Lambda|}
\frac{\Theta_1(\Lambda)}{\Theta_2(\Lambda)}\Big|
&\leq 
{\rm e}^{2\delta \partial|\Lambda|+\beta\hat{g}|\Lambda|}
\sum 
\prod_{j=1}^p
{\rm e}^{-\hat{\tau}(\beta)|\ga^1_j|}
{\rm e}^{-\beta\hat{g}|\ig^1_j|}\\
&\leq
{\rm e}^{2\delta \partial|\Lambda|+\beta\hat{g}|\Lambda|}
\sum 
\prod_{j=1}^p
{\rm e}^{-(\hat{\tau}(\beta)-C_0\delta)|\ga^1_j|}
\hat{\Theta}(\ig^1_j)\,.\nonumber
\end{align}
In the last line of \eqref{da} we interpret ${\rm e}^{-\beta\hat{g}|\ig^1|}$ as a partition function (up to a boundary term),
since by Lemma \ref{lem1.1},
$$
{\rm e}^{-\beta\hat{g}|\ig^1|}\leq \hat{\Theta}(\ig^1)\,
{\rm e}^{C_0\delta|\ga^1|}\,.
$$
We sum over external contours in \eqref{da} and get
$$
\Big|{\rm e}^{-\beta z(h(\psi_1)-h(\psi_2))|\Lambda|}
\frac{\Theta_1(\Lambda)}{\Theta_2(\Lambda)}\Big|
\leq 
{\rm e}^{2\delta\partial|\Lambda|+\beta\hat{g}|\Lambda|}
\hat{\Theta}(\Lambda)\leq
{\rm e}^{3\delta\partial|\Lambda|}\,.
$$

\end{proof}

It is not difficult to prove more regularity for the curve 
$\nu\mapsto\mu^*(\nu;\beta)$. We need below only the following result.

\begin{lemma}\label{lem2.0}
Let $0<\delta<1$.
If $\beta$ is sufficiently large, then for all $n\geq 1$
$\displaystyle\frac{d}{d\nu}\mu^*_n(0;\beta)=0$, and
$$
\big|\frac{d^2}{d\nu^2}\mu^*_n(\nu;\beta)\big|\leq  
\frac{2\delta}{\Delta-2\delta}\Big(\Big(\frac{2\delta}{\Delta-2\delta}\Big)^2
+\frac{2\delta}{\Delta-2\delta}+1\Big)  \,.
$$  
\end{lemma}

\begin{proof}
Let $\delta$ be as in the proof of Proposition \ref{pro2.1}.
Because the free energies $f_1^{n-1}$ and $f_2^{n-1}$ are real on the real axis, it follows that 
$\nu\mapsto\mu^*_n(\nu;\beta)$ is even, and therefore
$\displaystyle\frac{d}{d\nu}\mu^*_n(0;\beta)=0$. By definition
$\mu^*_n(\nu;\beta)$ is 
solution of 
$$
\mathrm{Re}\big(f_2^{n-1}(\mu_{n}^*(\nu;\beta)+i\nu)-
f_1^{n-1}(\mu_{n}^*(\nu;\beta)+i\nu)\big)=0\,,
$$
which implies that
$$
\Delta\frac{d\mu^*_n}{d\nu}=
\frac{d}{d\mu}\mathrm{Re}\big(g_1^{n-1}-g_2^{n-1}\big)\frac{d\mu^*_n}{d\nu}
+\frac{d}{d\nu}\mathrm{Re}\big(g_1^{n-1}-g_2^{n-1}\big)\,
$$
and
\begin{align*}
\Delta\frac{d^2\mu^*_n}{d\nu^2}
&=
\frac{d}{d\mu}\mathrm{Re}\big(g_1^{n-1}-g_2^{n-1}\big)\frac{d^2\mu^*_n}{d\nu^2}
+\frac{d^2}{d\mu^2}\mathrm{Re}\big(g_1^{n-1}-g_2^{n-1}\big)
\big(\frac{d\mu^*_n}{d\nu}\big)^2\\
&
+\frac{d^2}{d\mu d\nu}\mathrm{Re}\big(g_1^{n-1}-g_2^{n-1}\big)
\frac{d\mu^*_n}{d\nu}
+\frac{d^2}{d\nu^2}\mathrm{Re}\big(g_1^{n-1}-g_2^{n-1}\big)\,.
\end{align*}
From the proof of Proposition \ref{pro2.1} we have on $\U_m$,
$$
\big|\frac{d}{dz}\omega_m(\ga)\big|\leq 
\beta C_3{\rm e}^{-\tau_2(\beta)|\ga|}\,.
$$
Let $\tau_3(\beta):=\tau_1(\beta)-2\frac{d}{d-1}$.
A similar proof shows that for $\beta$ sufficiently large, there exists $C_4$
so that for any $m$
$$
\big|\frac{d^2}{dz^2}\omega_m(\ga)\big|\leq 
\beta^2 C_4{\rm e}^{-\tau_3(\beta)|\ga|}\,.
$$
Assume  that $\beta$ is large enough so that 
$$
\beta \max\{C_4,C_3^2\}K_2{\rm e}^{-\tau_3(\beta)|\ga|}\leq\delta\,.
$$
Then Lemma \ref{lem1.1} gives for $G^{n-1}:=\mathrm{Re}\big(g_1^{n-1}-g_2^{n-1}\big)$
$$
\big|\frac{d}{d\mu}G^{n-1}\big|\leq2\delta\;,\;
\big|\frac{d}{d\nu}G^{n-1}\big|\leq2\delta\,,
$$
$$
\big|\frac{d^2}{d\mu^2}G^{n-1}\big|\leq 2\delta\;,\;
\big|\frac{d^2}{d\nu^2}G^{n-1}\big|\leq 2\delta\;,\;
\big|\frac{d^2}{d\mu d\nu}G^{n-1}\big|\leq 2\delta\,.
$$
Hence
$$
\big| \frac{d\mu^*_n}{d\nu}\big|
\leq\frac{2\delta}{\Delta-2\delta}\;,\;
\big| \frac{d^2\mu^*_n}{d\nu^2}\big|
\leq\frac{2\delta}{\Delta-2\delta}\Big(\Big(\frac{2\delta}{\Delta-2\delta}\Big)^2
+\frac{2\delta}{\Delta-2\delta}+1\Big)\,.
$$

\end{proof}

\begin{proposition}\label{pro2.2}
Under the conditions of Proposition \ref{pro2.1},
there exist $\beta_0\in\R^+$ and $p\in\N$ so that the following holds for all $\beta\geq \beta_0$. Let 
$$
\tau^\prime(\beta):=\tau(\beta)-\max\big\{\frac{d}{d-1},p\big\}\,.
$$
A) If $\mu+i\nu\in\U_0$ and $\mu\leq \mu^*(\nu;\beta)$, then 
$$
\big|\frac{d}{dz}\omega(\ga^2)(z)\big|\leq
\beta C_3{\rm e}^{-\tau^\prime(\beta)|\ga^2|}\,.
$$
B) If $\mu+i\nu\in\U_0$ and $\mu\geq \mu^*(\nu;\beta)$, then 
$$
\big|\frac{d}{dz}\omega(\ga^1)(z)\big|\leq
\beta C_3{\rm e}^{-\tau^\prime(\beta)|\ga^1|}\,.
$$
\end{proposition}

\begin{proof}
We consider the iterative construction of the proof of Proposition
\ref{pro2.1} with the same choice of the sequences $\{b^q_n\}$. 
Suppose that $z=\mu+i\nu\in\U_{n-1}\backslash\U_n$ and $\mu\leq \mu^*(\nu;\beta)$.
Suppose that $V(\ga^2)\leq n$.
We get (see \eqref{derivative})
$$
\big|\frac{d}{dz}\omega(\ga^2)\big|\leq
\beta C_3|\ga^2|^\frac{d}{d-1}|\omega(\ga^2)|\,.
$$
Since by Proposition \ref{pro2.1} $\omega(\ga^2)$ is 
$\tau(\beta)$-stable, we get for all $\ga^2$ such that $V(\ga^2)\leq n$,
$$
\big|\frac{d}{dz}\omega(\ga^2)\big|\leq
\beta C_3|\ga^2|^\frac{d}{d-1}{\rm e}^{-\tau(\beta)|\ga^2|}
\leq \beta C_3{\rm e}^{-\tau^\prime(\beta)|\ga^2|}\,.
$$
Suppose that $V(\ga^2)\geq n+1$.
We estimate the derivative at $z$ of $\omega(\ga^2)$  using Cauchy's formula
with a circle of center $z$ contained in $\{\mu+i\nu:\mu\leq\mu^*(\nu;\beta)\}$. We estimate from below 
$|\mathrm{Re}z-\mu^*(\nu;\beta)|$ when $z\in\U_{n-1}\backslash\U_n$,
uniformly in $\nu$.
$$
|\mathrm{Re}z-\mu^*|\geq |\mathrm{Re}z-\mu_n^*|-|\mu^*_n-\mu^*|\geq b_n^2-|\mu^*_n-\mu^*|\,.
$$
We estimate $|\mu^*_n-\mu^*|$ by first estimating
$|\mu^*_m-\mu^*_n|$. Let $m>n$; then, since $\mu^*_m\in\U_n$,
\begin{align*}
0
&=
\mathrm{Re}\big(f^{m-1}_2(\mu^*_m)-f^{m-1}_1(\mu^*_m)\big)
-\mathrm{Re}\big(f^{n-1}_2(\mu^*_n)-f^{n-1}_1(\mu^*_n)\big)\\ 
&=
\mathrm{Re}\big(f^{m-1}_2(\mu^*_m)-f^{n-1}_2(\mu^*_m)\big)
-\mathrm{Re}\big(f^{m-1}_1(\mu^*_m)-f^{n-1}_1(\mu^*_m)\big)\\
&\quad+
\mathrm{Re}\big(f^{n-1}_2(\mu^*_m)-f^{n-1}_2(\mu^*_n)\big)
-\mathrm{Re}\big(f^{n-1}_1(\mu^*_m)
-f^{n-1}_1(\mu^*_n)\big)\,.
\end{align*}
From \eqref{estimate} we get
$$
|\mu^*_m(\nu;\beta)-\mu^*_n(\nu;\beta)|\leq
\frac{2\delta^{l(n)}}{\beta(\Delta-2\delta)}\quad\forall\;m>n\,,
$$
so that
\begin{equation}\label{distance}
|\mu^*(\nu;\beta)-\mu^*_n(\nu;\beta)|\leq
\frac{2\delta^{l(n)}}{\beta(\Delta-2\delta)}\,.
\end{equation}
If $V(\ga^2)\geq n+1$, then $|\ga^2|\geq l(n+1)$. 
Choose
$p\in\N$ so that for all $n\geq 1$
$$
\frac{1}{|\gd|^p}\leq\Big(\frac{1}{2dn^\frac{d-1}{d}}\Big)^p
\leq
\frac{\chi\theta^\prime}{(\Delta+2\delta)n^\frac{1}{d}}-
\frac{2\delta^{l(n)}}{\beta(\Delta-2\delta)}\leq b^2_n-|\mu^*-\mu^*_n|
\leq |\mathrm{Re}z-\mu^*|\,.
$$
We use Cauchy's formula
with a circle of center $z$ and radius $|\ga^2|^{-p}$ and get
$$
\big|\frac{d}{dz}\omega(\ga^2)\big|\leq 
|\ga^2|^p{\rm e}^{-\tau(\beta)|\ga^2|}\leq
{\rm e}^{-\tau^\prime(\beta)|\ga^2|}\,.
$$

\end{proof}

\subsection{Analytic continuation of the weights of contours at $\mu^*$}
\label{subsection2.2}

In this subsection
we consider how the weight $\omega(\ga^2)$ for a contour with boundary condition $\psi_2$ behaves as function of $z=\mu+i\nu$ in the vicinity
of $z^*:=\mu^*(\nu;\beta)+i\nu$. We improve the domains of analyticity
of the weights of contours, by making a new choice of the sequences
$\{b_n^q\}$, $q=1,2$. The main result of this subsection is Proposition \ref{pro2.3}.
At $z^*$ the (complex) free energies
$f_q$, $q=1,2$, are well-defined and can be computed by the cluster expansion
method.
Moreover, 
$$
\mathrm{Re}f_2(z^*)=\mathrm{Re}f_1(z^*)\,.
$$
Therefore
$$
\mathrm{Re}g_1(z^*)+\mu^*(\nu;\beta)h(\psi_1)=
\mathrm{Re}g_2(z^*)+\mu^*(\nu;\beta)h(\psi_2)\,.
$$
With $\delta$  as in the proof of Proposition \ref{pro2.1}, we get
$$
|\mu^*(\nu;\beta)|
\leq \frac{2\delta}{\beta\Delta}\,,
$$
and
$$
|\omega(\gq)(z^*)|\leq 
\exp\big[-\beta\|\gq\| +\frac{2C_1\delta}{\Delta}|\gq| 
+\delta C_0|\gq|\big]\,,\;\forall\;\gq\,.
$$
We set
$$
\mu^*:=\mu^*(0;\beta)\,,
$$
and adopt the following
convention: if a quantity, say $\cH$ or $f_q$, is evaluated at the transition point $\mu^*$, we simply write $\cH^*$ or $f_q^*$.

The analyticity properties of $\omega(\ga^2)$ near $\mu^*$ are controlled by 
isoperimetric inequalities 
\begin{equation}\label{iso2}
V(\ga^2)^{\frac{d-1}{d}}\leq \chi_2(n)^{-1}\|\ga^2\|\quad
\forall\;\ga^2\,,\,V(\ga^2)\geq n\,.
\end{equation}
The difference with \eqref{iso12} is that only contours with boundary condition $\psi_2$ and  $V(\ga^2)\geq n$ are considered for a given $n$.
By definition the isoperimetric constants $\chi_2(n)$ verify
$$
\chi_2(n)^{-1}:=\inf\Big\{C:\,\frac{V(\ga^2)^\frac{d-1}{d}}{\|\ga^2\|}\leq C\,,\,\forall\;
\ga^2\;\text{such that}\;
V(\ga^2)\geq n\Big\}\,.
$$
$\chi_2(n)$ is a bounded increasing sequence; we set $\chi_2(\infty):=\lim_{n}\chi_2(n)$, and  define
$$
R_2(n):=\inf_{m:m\leq n}\frac{\chi_2(m)}{m^\frac{1}{d}}\,.
$$
There are similar definitions for $\chi_1(n)$ and $R_1(n)$. The corresponding
isoperimetric inequalities control the analyticity properties of 
$\omega(\ga^1)$ around $\mu^*$.

\begin{lemma}\label{lem2.1}
For any $\chi_q^\prime<\chi_q(\infty)$, there exists
$N(\chi_q^\prime)$ such that for all $n\geq N(\chi_q^\prime)$,
$$
\frac{\chi_q^\prime}{n^\frac{1}{d}}\leq R_q(n)\leq 
\frac{\chi_q(\infty)}{n^\frac{1}{d}}\,.
$$   
For $q=1,2$, $n^a\mapsto n R_q(n)$ is increasing in $n$, provided that
$a\geq \frac{1}{d}$.
\end{lemma}

\begin{proof}
Let $q=2$ and suppose that 
$$
R_2(n)=\frac{\chi_2(m)}{m^\frac{1}{d}}\quad\text{for $m<n$.}
$$
Then $R_2(m^\prime)=R_2(n)$ for all $m\leq m^\prime\leq n$.  Let $n^\prime$ be the largest $n\geq m$ such that
$$
R_2(n)=\frac{\chi_2(m)}{m^\frac{1}{d}}\,.
$$
We have $n^\prime<\infty$, otherwise
$$
0< R_2(m)=R_2(n)\leq \frac{\chi_2(\infty)}{n^\frac{1}{d}}\quad\forall\;n\geq m\,,
$$
which is impossible. Therefore, either 
$$
R_2(n^\prime)=\frac{\chi_2(n^\prime)}{{n^\prime}^\frac{1}{d}}\quad
\text{or}\quad R_2(n^\prime+1)=\frac{\chi_2(n^\prime+1)}
{(n^\prime+1)^\frac{1}{d}}\,,
$$
and for all $k\geq n^\prime+1$, since $\chi_2(m)$ is increasing,
\begin{equation}\label{infinitely}
R_2(k)=\inf_{m\leq k}\frac{\chi_2(m)}{m^\frac{1}{d}}
=\inf_{n^\prime\leq m\leq k}\frac{\chi_2(m)}{m^\frac{1}{d}}
\geq
\inf_{n^\prime\leq m\leq k}\frac{\chi_2(n^\prime)}{m^\frac{1}{d}}
=\frac{\chi_2(n^\prime)}{{k}^\frac{1}{d}}\,.
\end{equation}
Inequality \eqref{infinitely} is true for infinitely many $n^\prime$; since there exists $m$ such that $\chi_2^\prime\leq\chi_2(m)$, the first statement is proved.

On an interval of constancy of $R_2(n)$, 
$n\mapsto n^{a} R_2(n)$ is increasing. 
On the other hand, if on $[m_1,m_2]$
$$
R_2(n)=\frac{\chi_2(n)}{n^\frac{1}{d}}\,,
$$
then $n\mapsto n^a R_2(n)$ is increasing on $[m_1,m_2]$
since $n\mapsto \chi_2(n)$ and $n\mapsto n^{a-\frac{1}{d}}$
are increasing.

\end{proof}

The next proposition gives the domains of analyticity and the stability properties of the weights $\omega(\ga)$ needed for estimating the derivatives of the free energy.
 
\begin{proposition}\label{pro2.3}
Let $0<\theta<1$ and $0<\varepsilon<1$ so that $\rho(1-\theta)-\varepsilon>0$.\\
There exist $0<\delta<1$, $0<\theta^\prime<1$ and $\beta_0\in\R^+$, such that for all $\beta\geq\beta_0$ $\omega(\ga^2)$ is analytic and $\tau_1(\beta;\theta^\prime)$-stable in a complex neighborhood of 
$$
\big\{z\in\C:\,\mathrm{Re}z\leq \mu^*(\mathrm{Im}z;\beta)
+\theta\Delta^{-1}R_2(V(\ga^2))\big\}\cap\U_0\,.
$$
Moreover
$$
\big|\frac{d}{dz}\omega(\ga^2)\big|\leq\beta C_3{\rm e}^{-\tau_2(\beta;\theta^\prime)|\ga^2|}\,.
$$
Similar properties hold for $\omega(\ga^1)$ in a 
complex neighborhood of 
$$
\big\{z\in\C:\,\mu^*(\mathrm{Im}z;\beta)-\theta\Delta^{-1}R_1(V(\ga^1))\leq \mathrm{Re}z\big\}\cap\U_0\,.
$$
$\tau_1(\beta;\theta^\prime)$ and $\tau_2(\beta;\theta^\prime)$
are defined at \eqref{tau1} and \eqref{tau2}.
\end{proposition}

\begin{proof}
If in the iterative method of the proof of Proposition \ref{pro2.1} we find
$0<\theta^\prime<1$ and  $b^1_n$, $b_n^2$, so that 
\eqref{mc2}, \eqref{mc3} and
\begin{equation}\label{mc4}
\big(\mu^*(\nu;\beta)-\theta\Delta^{-1}R_1(n),
\mu^*(\nu;\beta)+\theta\Delta^{-1}R_2(n)\big)\subset U_n(\nu;\beta)
\end{equation}
hold, then Proposition \ref{pro2.3} is true. Formula
\eqref{mc4} is satisfied if (see \eqref{distance})
$$
b^q_n\geq \theta\Delta^{-1}R_q(n)+\frac{2\delta^{l(n)}}{\beta(\Delta-2\delta)}\,,
$$
and this is the case if 
$$
b_n^q:=\theta \Delta^{-1}R_q(n)+\frac{C}{\beta}\delta^{n^\frac{1}{4}}\,,
$$
with $C$ a suitable constant, which is chosen so that \eqref{mc1} is also satisfied. If $\beta$ is large enough and $\delta$ small enough, then there exists $\theta^\prime<1$ so that \eqref{mc2} and \eqref{mc3} hold. Indeed, let $V(\ga^2)=n$, $z=\mu+i\nu$ and
$\mu\leq \mu^*(\nu;\beta)+b_n^2$; then
\begin{align*}
-\mathrm{Re}\big(f_1^{n-1}(z)-f_2^{n-1}(z)\big)\frac{V(\ga^2)}{\|\ga^2\|}
&\leq
(\Delta+2\delta)b_n^2\frac{n^\frac{1}{d}}{\chi_2(n)}\\
&\leq
\frac{\Delta+2\delta}{\Delta}\theta +\frac{C}{\beta}
\delta^{n^\frac{1}{4}}\frac{n^\frac{1}{d}}{\chi_2(n)}\leq\theta^\prime\,.
\end{align*}
\end{proof}

\subsection{Derivatives of the free energy at finite volume}\label{subsection2.3}

Although non-analytic behavior of the free energy occurs only in the thermodynamical limit, most of the analysis is done at finite volume.  We write 
$$
[g]^{(k)}_{\;t^\prime}:= 
\left.\frac{\diff^k}{\diff t^k}g(t)\right|_{t=t^\prime}\, 
$$
for the $k^{\mathrm{th}}$ order derivative at $t^\prime$ of the function $g$. 
The method of Isakov \cite{I1} allows to get estimates of the derivatives of the free energy at $\mu^*$, which are \emph{uniform in  the volume}. We consider the case of the boundary condition $\psi_2$. The other case is similar.
We tacitly assume that $\beta$ is large enough so that Lemma \ref{lem1.1}
and all results of subsections \ref{subsection2.1} and \ref{subsection2.2} are valid.
The main tool for estimating the derivatives of the free energy is Cauchy's formula. However, we need to establish several results before we can obtain the desired estimates on the derivatives of the free energy.  The preparatory work is done in this subsection, which is divided into three subsections. In \ref{subsubsection2.3.1} we give an
expression of the derivatives of the free energy in terms of the derivatives of
a free energy of a contour $u(\gd)=-\log(1+\phi_\Lambda(\gd))\approx
-\phi_\Lambda(\gd)$ (see \eqref{ufi}). The main work is to estimate
$$
\frac{k!}{2\pi i}\oint_{\partial D_r}
\frac{\phi_\Lambda(\gd)^n(z)}{(z-\mu^*)^{k+1}}\,dz\,.
$$
The boundary of the disc $D_r$ is decomposed  naturally into two parts,
$\partial D_r^g$ and $\partial D_r^d$, and the integral into two integrals
$I_{k,n}^g(\gd)$ and $I_{k,n}^d(\gd)$ (see \eqref{g} and \eqref{d}).
In \ref{subsubsection2.3.2} we prove the upper bound 
\eqref{easy} for $I_{k,n}^g(\gd)$, and in \ref{subsubsection2.3.3} we evaluate $I_{k,n}^d(\gd)$ by the stationary phase method, see
\eqref{grand} and \eqref{constantes}. This is a key point
in the proof of Theorem \ref{thm1.1}, since we obtain lower and upper
bounds for $I_{k,n}^d(\gd)$. 

\subsubsection{An expression for the derivatives of the free energy}\label{subsubsection2.3.1}

Let $\Lambda=\Lambda(L)$ be the cubic box
$$
\Lambda(L):=\{z\in\Z^d:\,|x|\leq L\}\,.
$$
We introduce a linear order, denoted by $\leq$, among all contours $\gq\subset\Lambda$
with boundary condition $\psi_q$.
We assume that the linear order is such that
$V(\ga^{\prime q})\leq V(\gq)$ if $\ga^{\prime q}\leq \gq$.  There exists  a natural enumeration of the contours by the positive integers.
The predecessor of $\gq$ in that enumeration (if $\gq$ is not the smallest contour) is
denoted by $i(\gq)$.  
We introduce the restricted partition function 
$\Theta_{\gq}(\Lambda)$, which is
computed with the contours of  
\begin{equation}\nonumber 
\cC_\Lambda(\gq):= \{\ga^{\prime q} \subset\Lambda :\ga^{\prime q}\leq \gq\}\,,  
\end{equation}
that is 
\begin{equation}\label{toto}
\Theta_{\gq}(\Lambda):=1+\sum\prod_{i=1}^n\omega(\gq_i)\,,
\end{equation}
where the sum is over all families of compatible contours
$\{\gq_1,\ldots,\gq_n\}$ which belong to $\cC_\Lambda(\gq)$.
The partition function $\Theta_q(\Lambda)$ is written as a finite product
\begin{equation}\nonumber 
\Theta_q(\Lambda)= 
\prod_{\gq\subset\Lambda} 
\frac{\Theta_{\gq}(\Lambda)}{\Theta_{i(\gq)}(\Lambda)}\,.
\end{equation} 
By convention $\Theta_{i(\gq)}(\Lambda):=1$ when $\gq$ is the smallest contour. 
We set
$$
\uq:=-\log \frac{\Theta_{\gq}(\Lambda)}{\Theta_{i(\gq)}(\Lambda)}\,.
$$
$\uq$ is the free energy cost for introducing the new contour $\gq$ in the restricted model, where all contours verify ${\ga^\prime}^q\leq\gq$. We have the identity 
\begin{align*} 
\Theta_{\gq}(\Lambda)
&=
\Theta_{i(\gq)}(\Lambda)+ \omega(\gq)\,\Theta_{i(\gq)}(\Lambda(\gq))\\
&=
\Theta_{i(\gq)}(\Lambda)\left(1+\omega(\gq)\,
\frac{\Theta_{i(\gq)}(\Lambda(\gq))}{\Theta_{i(\gq)}(\Lambda)}\right)\,. 
\end{align*} 
In this last expression $\Theta_{i(\gq)}(\Lambda(\gq))$ denotes the restricted partition function 
$$ 
\Theta_{i(\gq)}(\Lambda(\gq)):= 1+\sum\prod_{i=1}^n\omega(\gq_i)\,,
$$
where the sum is over all families of compatible contours
$\{\gq_1,\ldots,\gq_n\}$ which belong to $\cC_\Lambda(i(\gq))$, and such that
$\{\gq,\gq_1,\ldots,\gq_n\}$ is a compatible family.
We also set  
\begin{equation}\nonumber
\phi_\Lambda(\gq):=\omega(\gq)\, 
\frac{\Theta_{i(\gq)}(\Lambda(\gq))}{\Theta_{i(\gq)}(\Lambda)}\,. 
\end{equation} 
With these notations 
\begin{equation}\label{ufi}
\uq=-\log\big(1+\phi_\Lambda(\gq)\big)=\sum_{n\geq 1}\frac{(-1)^n}{n}\phi_\Lambda(\gq)^n\,,
\end{equation}
and for $k\geq 2$
\begin{equation*}
|\Lambda|\beta [f^q_\Lambda]^{(k)}_{\,\mu^*}=\sum_{\gq\subset\Lambda}
[\uq]^{(k)}_{\,\mu^*}.
\end{equation*} 

We consider the case of the boundary condition $\psi_2$.
$[\phi_\Lambda(\gd)^n]^{(k)}_{\mu^*}$ is computed using Cauchy's formula,  
$$
[\phi_\Lambda(\gd)^n]^{(k)}_{\,\mu^*}=\frac{k!}{2\pi i}\oint_{\partial D_r}
\frac{\phi_\Lambda(\gd)^n(z)}{(z-\mu^*)^{k+1}}\,dz\,,
$$
where $\partial D_r$ is the boundary of a disc $D_r$ of radius $r$ and center
$\mu^*$ inside the analyticity region of Proposition \ref{pro2.3},
$$
\U_0\cap\big\{z\in\C:\,\mathrm{Re}z\leq
\mu^*(\mathrm{Im}(z);\beta)+\theta \Delta^{-1}R_2(V(\ga^2))\big\}\,.
$$
The function $z\mapsto \frac{\phi_\Lambda(\gd)^n(z)}{(z-\mu^*)^{k+1}}$ is real
on the real axis, so that 
$$
\overline{\frac{\phi_\Lambda(\gd)^n(\overline{z})}
{(\overline{z}-\mu^*)^{k+1}}}=
\frac{\phi_\Lambda(\gd)^n(z)}{(z-\mu^*)^{k+1}}\,,
$$
and consequently
\begin{align}\label{twoparts}
\frac{k!}{2\pi i}\oint_{\partial D_r}
\frac{\phi_\Lambda(\gd)^n(z)}{(z-\mu^*)^{k+1}}\,dz
&=
\mathrm{Re}\Big\{
\frac{k!}{2\pi i}\oint_{\partial D_r}
\frac{\phi_\Lambda(\gd)^n(z)}{(z-\mu^*)^{k+1}}\,dz
\Big\}\,.
\end{align}

\begin{remark}\label{remark}{\rm
From Lemma \ref{lem2.0},
there exists $C^\prime$ independent of $\nu$ and $n$, so that  
$$
\mu_n^*(\nu;\beta)\geq \mu^*_n(0;\beta)- C^\prime\nu^2\,.
$$
This implies that the region
$\{\mathrm{Re}z\leq
\mu^*-C^\prime(\mathrm{Im}z)^2+\theta\Delta^{-1} R_2(V(\gd))\}$ is always in the analyticity region of $\omega(\ga^2)$, which is given in Proposition \ref{pro2.3}. Therefore,
if
$$ 
C^\prime \leq\frac{1}{2\big(\theta\Delta^{-1} R_2(V(\gd))\big)^2}\,,
$$
then the disc $D_r$ of center $\mu^*$ and radius $r=\theta\Delta^{-1} R_2(V(\gd))$ is
inside the analyticity region of $\omega(\ga^2)$. This happens as soon as
$V(\ga^2)$ is large enough. }
\end{remark}

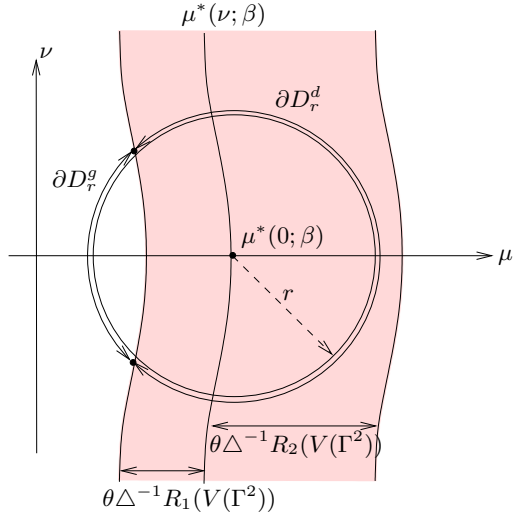
\begin{figure}[htbp]
\begin{center}
\input{integrale2.pstex_t}
\end{center}
\caption{{The decomposition of the integral into 
$I_{k,n}^g(\Gamma^2)$ and $I_{k,n}^d(\Gamma^2)$}}
\label{fig1}
\end{figure}

Assuming that the disc $D_r$ is inside  the analyticity region of $\omega(\ga^2)$, we decompose
$\partial D_r$ into 
$$
\partial D_r^g:=\partial D_r\cap\{z:\,\mathrm{Re}z\leq
\mu^*(\mathrm{Im}(z);\beta)-\theta \Delta^{-1}R_1(V(\ga^2))\}\,,
$$
and
$$
\partial D_r^d:=\partial D_r\cap\{z:\,\mathrm{Re}z\geq
\mu^*(\mathrm{Im}(z);\beta)-\theta \Delta^{-1}R_1(V(\ga^2))\}\,,
$$
and write \eqref{twoparts} as a sum of two integrals $I_{k,n}^g(\ga^2)$ and
$I_{k,n}^d(\ga^2)$ (see figure \ref{fig1}),
\begin{equation}\label{g}
I_{k,n}^g(\ga^2):=\mathrm{Re}\Big\{
\frac{k!}{2\pi i}\oint_{\partial D_r^g}
\frac{\phi_\Lambda(\gd)^n(z)}{(z-\mu^*)^{k+1}}\,dz
\Big\}
\end{equation}
and
\begin{equation}\label{d}
I_{k,n}^d(\ga^2):=\mathrm{Re}\Big\{
\frac{k!}{2\pi i}\oint_{\partial D_r^d}
\frac{\phi_\Lambda(\gd)^n(z)}{(z-\mu^*)^{k+1}}\,dz
\Big\}
\,.
\end{equation}

\subsubsection{An upper bound for $I_{k,n}^g(\ga^2)$}\label{subsubsection2.3.2} 
$I_{k,n}^g(\ga^2)$ is not the main contribution to \eqref{twoparts}, so that it is sufficient to get an upper bound for this integral.
Let $z\in\U_0$ and $\mathrm{Re}z\leq\mu^*\big(\mathrm{Im}(z);\beta\big)$.
We set
$$
\overline{\gd}:=\{x\in\Z^d:\,d(x,\sgd)\leq 1\}\,.
$$
There exists a constant $C_5$ such that $|\overline{\gd}|\leq C_5|\gd|$.
>From \eqref{ouf} we get
$$
|\omega(\gd)|\leq\exp\big[-\beta\|\gd\| +\beta |\mathrm{Re}z|C_1|\gd|
+3C_0\delta|\gd|\big]\,,
$$
and by the cluster expansion method
$$
\Big|\frac{\Theta_{i(\gd)}(\Lambda(\gd))}
{\Theta_{i(\gd)}(\Lambda)}\Big|
\leq
{\rm e}^{\delta|\overline{\gd}|}\leq {\rm e}^{\delta C_5|\gd|}\,.
$$
We set
$$
\zeta:=z-\mu^*\,.
$$
Therefore, there exists a constant $C_6$ so that 
\begin{equation*}
|\phi_\Lambda(\gd)|\leq {\rm e}^{-\beta\|\gd\|(1- C_6\delta
-|\mathrm{Re}\zeta|C_1\rho^{-1})}
\quad
\text{if}\quad
\mathrm{Re}\zeta\leq\mu^*\big(\mathrm{Im}(\zeta);\beta\big)-\mu^*\,.
\end{equation*}
This upper bound implies
\begin{equation}\label{easy}
I_{k,n}^g(\ga^2)\leq
\frac{k!}{r^k}\,
{\rm e}^{-n\beta \|\gd\|(1-C_6\delta -rC_1\rho^{-1})}\,.
\end{equation}

\subsubsection{Lower and upper bounds for $I_{k,n}^d(\ga^2)$}\label{subsubsection2.3.3}
In order to apply the stationary phase method to evaluate  $I_{k,n}^d(\ga^2)$,
we first rewrite $\phi_\Lambda(\gd)$ in the following form,
$$
\phi_\Lambda(\gd)(z)=\phi_\Lambda^*(\gd)\,{\rm e}^{\beta\Delta V(\gd)(\zeta+
\tg(\gd)(\zeta))}\,,
$$
where $\tg(\gd)$ is an analytic function of $\zeta$ in a neighborhood of $\zeta=0$
and $\tg(\gd)(0)=0$. 
Let
$$
\mu^*\big(\mathrm{Im}(z);\beta\big)-\theta \Delta^{-1}R_1(V(\ga^2))\leq \mathrm{Re}z\leq
\mu^*\big(\mathrm{Im}(z);\beta\big)+\theta \Delta^{-1}R_2(V(\ga^2))\,.
$$
In this region (see figure \ref{fig1}) we control the weights of contours with 
boundary conditions
$\psi_2$ and $\psi_1$, whose volume is smaller than $V(\Gamma^2)$. 
By the cluster expansion method there exists an analytic function $\tg(\gd)$, which is real on the real axis, so that 
\begin{align*}
\phi_\Lambda(\gd)
&=
\exp\Big[-\beta\cH(\vfi_{\gd}|\psi_2)+
\underbrace{\log\frac{\Theta_1(\igud)}{\Theta_2(\igud)}+
\log\frac{\Theta_{i(\gd)}(\Lambda(\gd))}{\Theta_{i(\gd)}(\Lambda)}}_{:=
\tG(\gd)(z)}\Big]\\
&=
\phi^*_\Lambda(\gd)
\exp\Big[\beta \Delta V(\gd)\zeta+
\underbrace{\int_{\mu^*}^{\mu^*+\zeta}
\Big(\frac{d}{dz}\tG(\gd)(z)-\beta a(\vfi_{\gd})\Big)dz}_{:=
\beta \Delta V(\gd)\tg(\gd)(\zeta)}\Big]\,.
\end{align*}
For large enough $\beta$, $\tau^\prime(\beta)\geq \tau_2(\beta;\theta^\prime)$, so that we get from Lemma \ref{lem1.1} and Propositions \ref{pro2.1} to \ref{pro2.3}
\begin{align}\label{neu}
\big|\frac{d}{d\zeta}\tg(\gd)(\zeta)\big|
&\leq
2C_3K{\rm e}^{-\tau_2(\beta;\theta^\prime)}
\Big(\frac{1}{\Delta}+
\frac{C_0|\gd|}{\Delta V(\gd)}+\frac{|\overline{\gd}|}{\Delta V(\gd)}
\Big)+\frac{C_1|\gd|}{\Delta V(\gd)} \nonumber\\
&\leq
C_7\,{\rm e}^{-\tau_2(\beta;\theta^\prime)}+
C_8\frac{|\gd|}{V(\gd)}
\,,
\end{align}
for suitable constants $C_7$ and $C_8$. Moreover, there exists a constant
$C_9$ so that
\begin{equation}\label{c8}
\exp\big[-\beta\|\gd\|(1+C_9\delta)]\leq
\phi^*_\Lambda(\gd)\leq
\exp[-\beta\|\gd\|(1-C_9\delta)]\,.
\end{equation}

Let 
$$
c(n):=n\beta\Delta V(\gd)\,.
$$
We parametrize $\partial D_r^d$ by $z:=\mu^*+r{\rm e}^{i\alpha}$,
$-\alpha_1\leq\alpha\leq\alpha_2$, $0<\alpha_i\leq\pi$.
\begin{align*}
I_{k,n}^d(\ga^2)
&=
 \frac{\phi^*_\Lambda(\gd)^n}{2\pi r^k}\int_{-\alpha_1}^{\alpha_2} 
{\rm e}^{c(n)r\cos\alpha +c(n)\mathrm{Re}\,\tg(\gd)(\zeta)}\, 
\big[\cos(\widetilde{\psi}(\alpha))\big]
\,d\alpha\,, 
\end{align*}
where 
\begin{equation}\nonumber 
\widetilde{\psi}(\alpha):=c(n)r\sin\alpha +c(n)\,\mathrm{Im}\,\tg(\gd)(\zeta)-k\alpha\,. 
\end{equation} 
We search for a stationary phase point $\zeta_{k,n}=r_{k,n}{\rm e}^{i\alpha_{k,n}}$
defined by the equations
$$
\frac{d}{d\alpha}\Big(c(n)r\cos\alpha +c(n)\mathrm{Re}\,
\tg(\gd)\big(r{\rm e}^{i\alpha}\big)\Big)=0
\quad\text{and}\quad
\frac{d}{d\alpha}\widetilde{\psi}(\alpha)=0\,.
$$
These equations are equivalent to the equations 
($\;^\prime$ denotes the derivative with respect to $\zeta$)
\begin{align*}
c(n)\sin\alpha\big(1+\mathrm{Re}\,\tg(\gd)^\prime(\zeta)\big)
+\cos\alpha \mathrm{Im}\,\tg(\gd)^\prime(\zeta)
&=0\,;\\
c(n)r\cos\alpha\big(1+\mathrm{Re}\,\tg(\gd)^\prime(\zeta)\big)-r\sin\alpha
\mathrm{Im}\,\tg(\gd)^\prime(\zeta)
&=k\,.
\end{align*}
Since $\tg(\gd)$ is real on the real axis,  $\alpha_{k,n}=0$ and $r_{k,n}$ is solution of 
\begin{equation}\label{equationr}
c(n)r\big(1+\tg(\gd)^\prime(r)\big)=k\,.
\end{equation}

\begin{lemma}\label{lem2.3}
Let $\alpha_i\geq\pi/4$, $i=1,2$, $A\leq 1/25$ and 
$c(n)\geq 1$. If $\tg(\zeta)$ is analytic
in $\zeta$ in the disc $\{\zeta:\,|\zeta|\leq R\}$, 
real on the real axis, and for all $\zeta$
in that disc
$$
\big|\frac{d}{d\zeta}\tg(\gd)(\zeta)\big|\leq A\,,
$$
then there exists $k_0(A)\in\N$, such that for all integers $k$, 
$$
k\in\big[k_0(A),c(n)(1-2\sqrt{A})R \big]\,,
$$
there is a unique solution $0<r_{k,n}<R$
of \eqref{equationr}. Moreover,
\begin{align*}
\frac{{\rm e}^{cr_{k,n}+c(n)\,\tg(\gd)(r_{k,n})}}{10\sqrt{c(n)r_{k,n}}}
&\leq
\frac{1}{2\pi }\int_{-\alpha_1}^{\alpha_2}
{\rm e}^{c(n)r\cos\alpha +c(n)\mathrm{Re}\,\tg(\gd)}\, 
\big[\cos(\widetilde{\psi}(\alpha))\big]
\,d\alpha\\
&\leq 
\frac{{\rm e}^{c(n)r_{k,n}+c(n)\,\tg(\gd)(r_{k,n})}}{\sqrt{c(n)r_{k,n}}}\,.
\end{align*}
\end{lemma}

\begin{proof}
Existence and uniqueness of $r_{k,n}$ is a consequence of the monotonicity
of $r\mapsto c(n)r\big(1+\tg(\gd)^\prime(r)\big)$. The last part of Lemma
\ref{lem2.3} is proven in appendix of \cite{I1}.
The computation is relatively long, but standard. 

\end{proof}

Setting $c(n)=n\beta\Delta V(\gd)$ and $R=\theta\Delta^{-1}R_2(V(\gd))$ in
Lemma \ref{lem2.3} we get sufficient conditions for the existence of a stationary phase point and the following evaluation of the integral $I_{k,n}^d(\ga^2)$ by that method.
Since $r_{k,n}$ is solution of \eqref{equationr}, we have  
$$
k-\frac{kA}{(1+A)}=
\frac{k}{(1+A)}\leq c(n)r_{k,n}\leq \frac{k}{(1-A)}=k+\frac{kA}{(1-A)}\,,
$$
and
$$
c(n)|\tg(\gd)(r_{k,n})|=c(n)\,\Big|\int_{0}^{r_{k,n}}
\tg(\gd)^\prime(\zeta)d\zeta\Big|\leq A c(n)r_{k,n}\leq k\frac{A}{1-A}\,.
$$
Therefore  Lemma \ref{lem2.3} implies
\begin{align}\label{grand}
\frac{\sqrt{1-A}}{10\sqrt{k}}c_-^k\,c(n)^k\,\frac{k!\,{\rm e}^k}{k^k}
\,\phi^*_\Lambda(\gd)^n
&\leq
I_{k,n}^d(\ga^2) \\
&\leq
\frac{\sqrt{1+A}}{\sqrt{k}}c_+^k\,c(n)^k\,\frac{k!\,{\rm e}^k}{k^k}
\,\phi^*_\Lambda(\gd)^n\,,\nonumber
\end{align}
with
\begin{equation}\label{constantes}
c_\pm(A):=(1\pm A)\exp\Big[\pm\frac{2A}{1-A^2}\Big]\,.
\end{equation}
If $A$ converges to $0$, then $c_{\pm}$ converges to $1$. 
We assume that (see \eqref{neu}) 
\begin{equation}\label{2301}
C_7\,{\rm e}^{-\tau_2(\beta;\theta^\prime)}\leq\frac{A}{2}\quad
\text{and}\quad
C_8\frac{|\gd|}{V(\gd)}\leq\frac{A}{2}\,.
\end{equation}
$A$ can be chosen as small as we wish, provided that 
$\beta$ is large enough and $\frac{|\gd|}{V(\gd)}$ small enough. 

\subsection{Lower bounds on the derivatives of the free energy at finite volume}\label{subsection2.4}

We estimate the derivative of $[f^2_\Lambda]^{(k)}_{\mu^*}$ for large enough $k$. The main result of this subsection is Proposition \ref{pro2.5}.

Let $0<\theta<1$, $A\leq 1/25$, and set
$$
\hat{\theta}:=\theta(1-2\sqrt{A})\,.
$$
Let $\varepsilon^\prime>0$ and $\chi_2^\prime$ so that
\begin{equation}\label{chi2}
(1+\varepsilon^\prime)\chi_2^\prime> \chi_2(\infty)\,.
\end{equation}
The whole analysis depends on the parameters $\theta$ and $\varepsilon^\prime$. We fix the values of $\theta$, and
$\varepsilon^\prime$ by the following conditions,
which are needed for the proof of Proposition \ref{pro2.5}. We choose
$0<A_0<1/25$, $\theta$ and $\varepsilon^\prime$ so that
\begin{equation}\label{choice}
{\rm e}^\frac{1}{d}\,\frac{1}{\theta(1-2 \sqrt{A_0})}<
\frac{d}{d-1}\,\frac{c_{-}(A_0)^\frac{d-1}{d}}{1+\varepsilon^\prime}\quad
\text{and}\quad
\frac{1-2\sqrt{A_0}}{1+\varepsilon^\prime}\,\frac{d}{d-1}>1\,.
\end{equation}
This is possible, since
$$ 
\frac{d}{(d-1)\,{\rm e}^\frac{1}{d}}>1\,. 
$$  
Indeed,
\begin{align*} 
d\Big({\rm e}^{{\frac{1}{d}}}-1\Big) 
&= 
d\Big({\rm e}^{{\frac{1}{d}}}-1-\frac{1}{d}+\frac{1}{d}\Big) 
= 
\sum_{n\geq 2}\frac{1}{n!}\Big(\frac{1}{d}\Big)^{n-1}+1 \\
&= 
1+\sum_{n\geq 1}\frac{1}{(n+1)!}\Big(\frac{1}{d}\Big)^{n}\\ 
&< 
1-\frac{1}{2d}+\sum_{n\geq 1}\frac{1}{n!}\Big(\frac{1}{d}\Big)^{n}={\rm e}^{{\frac{1}{d}}} -\frac{1}{2d}
\,.
\end{align*} 
Notice that conditions \eqref{choice} are still verified with the same values of $\theta$ and $\varepsilon^\prime$ if we replace
$A_0$ by $0<A<A_0$. 
Given $\theta$, the value of $\theta^\prime$ is fixed in Proposition \ref{pro2.3}.
>From now we assume  that $\beta$ is so large that all results of subsections \ref{subsection2.1}
and \ref{subsection2.2} are valid. 
The value of $0<A<A_0$ is fixed in the proof of Lemma \ref{lem2.40}.

Given $k$ large enough, there is a natural distinction between contours
$\gd$ such that  $\hat{\theta}\beta V(\gd)R_2(V(\gd))\leq k$ and those
such that $\hat{\theta}\beta V(\gd)R_2(V(\gd))> k$. For the latter we 
can estimate $I_{k,n}^d(\gd)$ by the stationary phase method. 
We need as a matter of fact a finer distinction between contours.
We distinguish three classes of contours:
\begin{enumerate}
\item
$k$-small contours: $\hat{\theta}\beta V(\gd)R_2(V(\gd))\leq k$;
\item
fat contours: for $\eta\geq 0$, fixed later by \eqref{ep},
$V(\gd)^\frac{d-1}{d}\leq \eta\,\|\gd\|$;
\item
$k$-large and thin contours: $\hat{\theta}\beta V(\gd)R_2(V(\gd))>k$,
$V(\gd)^\frac{d-1}{d}> \eta \,\|\gd\|$.
\end{enumerate}
We make precise the meaning of \emph{$k$ large enough}.
By Lemma \ref{lem2.1} 
$V\mapsto VR_2(V)$ is increasing in $V$, and there exists $N(\chi_2^\prime)$
such that 
$$
R_2(V)\geq \frac{\chi_2^\prime}{ V^{\frac{1}{d}}}\quad\text{if}\quad
V\geq N(\chi_2^\prime)\,. 
$$
We assume that there is a $k$-small contour $\gd$ such that 
$V(\gd)\geq N(\chi_2^\prime)$, and that the maximal volume of the $k$-small contours is so large that remark \ref{remark} is valid. We also assume (see Lemma \ref{lem2.3})
that $k>k_0(A)$ and that for a $k$-large and thin contour (see \eqref{neu}
and \eqref{2301})
$$
C_8\frac{|\ga^2|}{V(\ga^2)}\leq \frac{C_8}{\rho\eta V(\ga^2)^\frac{1}{d}}\leq\frac{A}{2}\,,
$$
so that $|\tg(\gd)^\prime|\leq A$,
and 
\begin{equation}\label{petit}
\frac{ C_1 k}{\rho\Delta(1-A_0)\eta V(\gd)^\frac{1}{d}}
\leq \frac{k}{10}
\end{equation}
are verified. There exists $K(A,\eta,\beta)$ such that
if $k\geq K(A,\eta,\beta)$, then $k$ is \emph{large enough}.
>From now on $k\geq K(A,\eta,\beta)$.

\subsubsection{Contribution to $[f^q_\Lambda]^{(k)}_{\mu^*}$ from the
$k$-small and fat contours}

Let $\gd$ be a $k$-small contour. Since $V\mapsto R_2(V)$ is decreasing in 
$V$, $\ud$ is analytic in the region 
$$
\{z:\,\mathrm{Re}z\leq \mu^*(\mathrm{Im}z;\beta) +\theta\Delta^{-1}R_2(V^*)\}
\cap \U_0\,,
$$
where $V^*$ is the maximal volume of $k$-small contours. $V^*$ satisfies 
$$
{V^*}^\frac{d-1}{d}\leq \frac{k}{\hat{\theta}\beta \chi^\prime_2}\,.
$$
Hence
$$
\theta\Delta^{-1} R_2(V^*)\geq
\hat{\theta}\Delta^{-1}\chi^\prime_2 {V^*}^{-\frac{1}{d}}
\geq
\Delta^{-1}\big(\hat{\theta}\chi_2^\prime\big)^{\frac{d}{d-1}}
\beta^\frac{1}{d-1}k^{-\frac{1}{d-1}}\,.
$$
Since remark \ref{remark} is valid, we estimate the derivative of $\ud$ by Cauchy's formula with a disc centered at $\mu^*$ with radius 
$\Delta^{-1}\big(\hat{\theta}\chi_2^\prime\big)^{\frac{d}{d-1}}
\beta^\frac{1}{d-1}k^{-\frac{1}{d-1}}$.
There exists a constant $C_{10}$ such that
\begin{align}\label{petitscontours}
\Big|\sum_{\substack{\gd:\ig^2\ni 0\\ 
V(\gd)^\frac{d-1}{d}\leq \frac{k}{\hat{\theta}\beta \chi^\prime_2}}}
[\ud]^{(k)}_{\,\mu^*}\Big|
&\leq
C_{10}
\Big(\frac{\Delta}{\beta^\frac{1}{d-1}(\hat{\theta}\chi_2^\prime)^\frac{d}{d-1}}\Big)^k
k!\,k^\frac{k}{d-1}\,.
\end{align}

Let $\gd$ be a fat contour, which is not $k$-small. We use in Cauchy's formula a disc centered at $\mu^*$ with radius
$$
\hat{\theta}\Delta^{-1}\chi_2(1)V(\gd)^{-\frac{1}{d}}\leq
\theta \Delta^{-1} R_2(V(\gd))\,.
$$
We get (see \eqref{C2})
\begin{align*}
\big|[\phi_\Lambda(\gd)^n]^{(k)}_{\,\mu^*}\big|
&\leq 
k!\,
\left(\frac{\Delta V(\gd)^{\frac{1}{d}}}{\chi_2(1)\hat{\theta}}\right)^k
\,{\rm e}^{- n[\tau_1(\beta;\theta^\prime)-C_5\delta]|\gd|}\\
&\leq
k!\,
\left(\frac{\Delta\,(C_2\eta)^{\frac{1}{d-1}}}{\chi_2(1)\hat{\theta}}\right)^k
\,|\gd|^{\frac{k}{d-1}}
{\rm e}^{- n[\tau_1(\beta;\theta^\prime)-C_5\delta]|\gd|}\,.
\end{align*}
We  sum over $n$ and over $\gd$ using the inequality
$$
\sum_{m\geq 1}m^p\,{\rm e}^{-qm}\leq \frac{1}{q^p}\,\Gamma(p+1)\quad
(p\geq2\,,\,q\geq2)\,.
$$
There exist $C_{11}$ and $C_{12}(\theta^\prime)>0$ so that 
\begin{align*}
\sum_{\substack{\gd:\ig^d\ni 0\\ 
V(\gd)^\frac{d-1}{d}\leq \eta\|\gd\|\\
\gd\;\text{not $k$-small}}}
\big|[\ud]^{(k)}_{\,\mu^*}\big|
&\leq 
C_{11}
\left(\frac{\Delta\,(C_2\eta)^{\frac{1}{d-1}}}
{(C_{12}\beta)^\frac{1}{d-1}\chi_2(1)\hat{\theta}}\right)^k
k!\,\Gamma\Big(\frac{k}{d-1}+1\Big)\\
&\leq
C_{11}
\left(\frac{\Delta\,(C_2\eta)^{\frac{1}{d-1}}}
{(C_{12}\beta)^\frac{1}{d-1}\chi_2(1)\hat{\theta}}\right)^k
k!\,\,k^\frac{k}{d-1}\,.
\end{align*}
We choose $\eta$ so small that (see \eqref{petitscontours})
\begin{equation}\label{ep}
\frac{\Delta\,(C_2\eta)^{\frac{1}{d-1}}}
{(C_{12}\beta)^\frac{1}{d-1}\chi_2(1)\hat{\theta}}<
\frac{\Delta}{\beta^\frac{1}{d-1}(\hat{\theta}\chi_2(\infty))^\frac{d}{d-1}}
<
\frac{\Delta}{\beta^\frac{1}{d-1}(\hat{\theta}\chi_2^\prime)^\frac{d}{d-1}}\,.
\end{equation}

\subsubsection{Contribution to $[f^q_\Lambda]^{(k)}_{\mu^*}$ from the
$k$-large and thin contours}

For $k$-large and thin contours we get lower and upper bounds for 
$[\phi_\Lambda(\gd)^n]^{(k)}_{\,\mu^*}$. There are two cases.\\

\noindent
A. Assume that $R_1(V(\gd))\geq R_2(V(\gd))$, or that $V(\gd)$ is so large
that 
$$
\hat{\theta}\beta V(\gd)R_1(V(\gd))>k\,.
$$
For each $n\geq 1$ let $c(n)=n\beta\Delta V(\gd)$. Under these conditions we
can apply Lemma \ref{lem2.3} with a disc $D_{r_{k,n}}$
so that $\partial D_{r_{k,n}}=\partial D^d_{r_{k,n}}$. Indeed, if
$R_1(V(\gd))\geq R_2(V(\gd))$, then we apply Lemma \ref{lem2.3} with
$R=\theta \Delta^{-1} R_2(V(\gd))$, and in the other case we set
$R=\theta \Delta^{-1} R_1(V(\gd))$. In both cases $r_{k,n}<R$, which implies
$\partial D_{r_{k,n}}=\partial D^d_{r_{k,n}}$.
Therefore we get for $I_{k,n}^d(\gd)$
the lower and upper bounds \eqref{grand}.

\begin{lemma}\label{lem2.4}
There exists a function $D(k)$, $\lim_{k\ra\infty}D(k)=0$,
such that for $\beta$ sufficiently large and $A$ sufficiently small 
the following holds.
If $k\geq K(A,\eta,\beta)$ and $R_1(V(\gd))\geq R_2(V(\gd))$ or
$\hat{\theta}\beta V(\gd)R_1(V(\gd))>k$, then 
\begin{align*}
(1-D(k))\,[\phi_\Lambda(\gd)]^{(k)}_{\,\mu^*}
\leq
-[u_\Lambda(\gd)]^{(k)}_{\,\mu^*}
\leq
(1+D(k))\,[\phi_\Lambda(\gd)]^{(k)}_{\,\mu^*}\,.
\end{align*}
\end{lemma}

\begin{proof}
We have
$$
-[u_\Lambda(\gd)]^{(k)}_{\,\mu^*}
=[\phi_\Lambda(\gd)]^{(k)}_{\,\mu^*}+
[\phi_\Lambda(\gd)]^{(k)}_{\,\mu^*}\sum_{n\geq2}\frac{(-1)^{(n-1)}}{n}
\frac{[\phi_\Lambda(\gd)^n]^{(k)}_{\,\mu^*}}
{[\phi_\Lambda(\gd)]^{(k)}_{\,\mu^*}}\,.
$$
From \eqref{grand} there exists a constant $C_{13}$,
$$
\frac{[\phi_\Lambda(\gd)^n]^{(k)}_{\,\mu^*}}{[\phi_\Lambda(\gd)]^{(k)}_{\,\mu^*}}
\leq
C_{13}\,\phi_\Lambda^*(\gd)^{(n-1)}
\,\Big(\frac{c_+}{c_-}\Big)^k\,n^k\,.
$$
The isoperimetric inequality \eqref{iso2}, $R_2(n)\leq\chi_2(n)n^{-\frac{1}{d}}$ and the definition of  $k$-large volume contour imply
\begin{align*}
\beta\|\gd\|\geq\beta\chi_2(V(\gd))V(\gd)^\frac{d-1}{d}\geq
\hat{\theta}\beta R_2(V(\gd))V(\gd)\geq k\,.
\end{align*}
Let $b:=C_{9}\delta$ (see \eqref{c8}); we may assume 
$\frac{9}{10}-b\geq\frac{4}{5}$ by taking $\beta$ large enough. 
Then
\begin{align*}
\frac{c_+^k}{c_-^k}\sum_{n\geq 2}n^{k-1}{\rm e}^{-(n-1)(1-b)k}
&\leq
\frac{c_+^k}{c_-^k}\sum_{n\geq 2}
{\rm e}^{-\frac{1}{10}(n-1)k}{\rm e}^{-k\big[(\frac{9}{10}-b)(n-1)-\ln n\big]}\\
&\leq
\frac{c_+^k}{c_-^k}\sum_{n\geq 2}
{\rm e}^{-\frac{1}{10}(n-1)k}{\rm e}^{-k\big[\frac{4}{5}(n-1)-\ln n\big]}\\
&\leq
\Big(\frac{c_+}{c_-}{\rm e}^{-\frac{1}{10}}\Big)^k\,
\sum_{n\geq 1}{\rm e}^{-\frac{1}{10}nk}\,.
\end{align*}
We choose $A$ so small that 
$c_+(A){c_-(A)}^{-1}\,{\rm e}^{-\frac{1}{10}}\leq 1$.

\end{proof}

\noindent
B. The second case is when
$$
\hat{\theta}\beta V(\gd)R_1(V(\gd))\leq k\leq \hat{\theta}\beta V(\gd)R_2(V(\gd))\,.
$$
Since the contours are also thin,
\begin{align*}
\beta\|\gd\|
&\leq \eta^{-1}\hat{\theta}^{-1}\chi_1(1)^{-1}\beta\hat{\theta}\chi_1(1) V(\gd)^\frac{d-1}{d}\\
&\leq
\eta^{-1}\hat{\theta}^{-1}\chi_1(1)^{-1}\beta \hat{\theta}V(\gd)R_1(V(\gd))\\
&\leq 
\eta^{-1}\hat{\theta}^{-1}\chi_1(1)^{-1} k\equiv \lambda k\,.
\end{align*}
We choose $R=\beta\Delta^{-1}R_2(V(\gd))$ in Lemma \ref{lem2.3}. The integration in \eqref{twoparts} is decomposed into two parts (see figure \ref{fig1}).
We show that
the contribution from the 
integration over $\partial D^g_{r_{k,n}}$ is negligible for large enough $\beta$. Since $k\geq K(A,\eta,\beta)$ and the contours verify
$V(\gd)^\frac{d-1}{d}>\eta\|\gd\|$, we have 
\begin{equation*}
n\beta\|\gd\|r_{k,n}\leq \frac{k}{\Delta(1-A)\eta V(\gd)^\frac{1}{d}}
\leq 
\frac{ k}{\Delta(1-A_0)\eta V(\gd)^\frac{1}{d}}\,.
\end{equation*}
By definition of $K(A,\eta,\beta)$ (see \eqref{petit})
\begin{equation*}
n\beta\|\gd\|\rho^{-1} C_1r_{k,n}
\leq \frac{k}{10}\,.
\end{equation*}
From \eqref{easy} with $r=r_{k,n}$ we obtain that the contribution to
$|[u_\Lambda(\gq)]^{(k)}_{\,\mu^*}|$ is at most
$$
(1+A)^k\big(\beta\Delta V(\gd)\big)^k\exp\big(\frac{k}{10}\big)
\frac{k!}{k^k}\sum_{n\geq 1}n^k
{\rm e}^{-n\beta\|\gd\|(1-C_{6}\delta)}\,.
$$
As in the proof of Lemma \ref{lem2.4}, we 
choose $\beta$ large enough so that we can assume that $\frac{9}{10}-C_{6}\delta \geq\frac{4}{5}$. Then
\begin{align*}
\sum_{n\geq 1}n^k
{\rm e}^{-n\beta\|\gd\|(1-C_{6}\delta)}
&\leq
{\rm e}^{-\beta\|\gd\|(1-C_{6}\delta)}
\Big(1 +\sum_{n\geq 2}
{\rm e}^{-\frac{1}{10}(n-1)k}
{\rm e}^{-k\big[\frac{4}{5}(n-1)-\ln n\big]}\Big)\\
&\leq
{\rm e}^{-\beta\|\gd\|(1-C_{6}\delta)}
\Big(1 +\sum_{n\geq 1}
{\rm e}^{-\frac{1}{10}nk}\Big)\\
&=
{\rm e}^{-\beta\|\gd\|(1-C_{6}\delta)}\big(1+D(k)\big)
\,.
\end{align*}
Since $\beta\|\gd\|\leq\lambda k$, by choosing 
$A$ small enough and $\beta$ large enough, so that $\delta$ is small enough, we have
$$
(1-D(k))c_-^k{\rm e}^k{\rm e}^{-\beta\|\gd\|C_{9}\delta}
\geq
(1-D(k))c_-^k{\rm e}^k{\rm e}^{-k\lambda C_{9}\delta}
>
{\rm e}^{\frac{2k}{3}}\,
$$
and
$$
(1+D(k))(1+A)^k{\rm e}^\frac{k}{10}{\rm e}^{\beta\|\gd\|C_{6}\delta}
\leq 
(1+D(k))(1+A)^k{\rm e}^\frac{k}{10}{\rm e}^{\lambda kC_{6}\delta}
<
{\rm e}^{\frac{k}{3}}\,.
$$
If these inequalities are verified, then 
the contribution to $-[u_\Lambda(\gq)]^{(k)}_{\,\mu^*}$ coming from the 
integrations over $\partial D^g_{r_{k,n}}$ is negligible with respect to that coming from the 
integrations over $\partial D^d_{r_{k,n}}$. Taking into account
\eqref{grand} we get  
Lemma \ref{lem2.40}.

\begin{lemma}\label{lem2.40}
There exists $0<A^\prime\leq A_0$ so that for all 
$\beta$ sufficiently large, the following holds.
If $k\geq K(A^\prime,\eta,\beta)$ and $\gd$ is a $k$-large and thin contour,
then 
$$
-[u_\Lambda(\gd)]^{(k)}_{\,\mu^*}
\geq 
\frac{1}{20}(1-D(k))
\big(\beta \Delta V(\gd)\big)^k c_-^k\,\phi^*_\Lambda(\gd)\,.
$$
\end{lemma}

\begin{proposition}\label{pro2.5}
There exists $\beta^\prime$ so that for all 
$\beta>\beta^\prime$, the following holds.
There exists an increasing diverging sequence $\{k_n\}$ such that
for each $k_n$ there exists $\Lambda(L_n)$ such that 
for all $\Lambda\supset\Lambda(L_n)$
$$
-[f_\Lambda^2]^{(k_n)}_{\,\mu^*}\geq 
C_{14}^{k_n}\,k_n!^\frac{d}{d-1}\,
\Delta^{k_n}\beta^{-\frac{k_n}{d-1}}\,{\chi_2^\prime}^{-\frac{dk_n}{d-1}}
\,.
$$ 
$C_{14}>0$ is a constant independent of $\beta$, $k_n$ and $\Lambda$.
\end{proposition}

\begin{proof}
We compare the contribution of the small and fat contours with that of the large and thin contours for $k\geq K(A^\prime,\eta,\beta)$. The contribution of the small contours to $|[f_\Lambda^2]^{(k)}_{\,\mu^*}|$ is at most
$$
C_{10}\,\Delta^k\,\beta^{-\frac{k}{d-1}}\,(\hat{\theta}\chi_2^\prime)^{-\frac{kd}{d-1}}\,
k!\,k^\frac{k}{d-1}
\leq
C_{10}\,\Delta^k\,\beta^{-\frac{k}{d-1}}
\Big
(\frac{{\rm e}^\frac{1}{d}}{\hat{\theta}\chi_2^\prime}
\Big)^{k\frac{d}{d-1}}
{k!}^\frac{d}{d-1}\,.
$$
The contribution of the fat contours is much smaller by our choice of $\eta$
(see \eqref{ep}).
The contribution to $-[f_\Lambda^2]^{(k)}_{\,\mu^*}$ of each large and thin contour is nonnegative.  
By assumption \eqref{chi2} and the definition of the isoperimetric constant $\chi_2$,
there exists a sequence $\gd_n$, $n\geq 1$,  such that
$$
\lim_{n\ra\infty}\|\gd_n\|\ra \infty
\quad
\text{and} 
\quad
V(\gd_n)^{\frac{d-1}{d}}\geq \frac{\|\gd_n\|}{(1+\varepsilon^\prime)\chi^\prime_2}\,.
$$
Since $x^{k\frac{d}{d-1}}{\rm e}^{-x}$ has its maximum at $x=k\frac{d}{d-1}$, we set
$$
k_n:=\left\lfloor\frac{d-1}{d}\beta\|\gd_n\|\right\rfloor\,.
$$ 
For any $n$, $\gd_n$ is a thin and $k_n$-large volume contour, since by \eqref{choice}
\begin{align*}
\beta\,(1-2\sqrt{A^\prime})V(\gd)R_2(V(\gd))
&\geq
\beta\,(1-2\sqrt{A^\prime})V(\gd)^{\frac{d-1}{d}}\chi_2^{\prime}\\
&\geq
\frac{(1-2\sqrt{A^\prime})}{1+\varepsilon^\prime}\beta\|\gd_n\|
\geq 
k_n\,.
\end{align*} 
If $\Lambda\supset \gd_n$, then
\begin{align*}
-[u_\Lambda(\gd_n)]^{(k_n)}_{\,\mu^*}
&\geq
\frac{1-D(k)}{20}
\,\big[\beta \Delta c_- V(\gd_n)\big]^{k_n}\,\phi_\Lambda^*(\gd_n)\\
&\geq
\frac{1-D(k)}{20}\Delta^{k_n}\beta^{-\frac{k_n}{d-1}}
\Big(\frac{d\,c_-^\frac{d-1}{d}}
{(d-1)(1+\varepsilon^\prime)\chi_2^\prime}\Big)^{\frac{dk_n}{d-1}}
k_n^\frac{k_nd}{d-1}\phi_\Lambda^*(\gd_n)\,
\end{align*}
and (see \eqref{c8})
\begin{align*}
k_n^\frac{k_nd}{d-1}\phi_\Lambda^*(\gd_n)
&\geq
k_n^\frac{k_nd}{d-1}\exp\big[-\big(k_n\frac{d}{d-1}+1\big)(1+C_9\delta)\big]\\
&\sim
{k_n!}^\frac{d}{d-1}{\rm e}^{-C_9\delta\frac{d}{d-1}k_n}
\frac{{\rm e}^{-1-C_9\delta}}{(2\pi k_n)^\frac{d}{2(d-1)}}\,.
\end{align*}
By the choice \eqref{choice} of the parameters $\theta$ and $\varepsilon^\prime$, if $\delta$ is small enough, i.e. $\beta$ large enough, then 
$$
\frac{{\rm e}^\frac{1}{d}}{\theta(1-2\sqrt{A^\prime})}<
\frac{d}{d-1}\,\frac{c_-^\frac{d-1}{d}}{1+\varepsilon^\prime}
{\rm e}^{-C_9\delta}\,.
$$
Hence the contributions 
of the small and fat contours are negligible for large $k_n$ (see 
\eqref{petitscontours} and \eqref{ep}).
Let $\Lambda(L_n)$ be a box which contains at 
least $|\Lambda(L_n)|/4$ translates of $\gd_n$.
For any $\Lambda\supset\Lambda(L_n)$, if $k_n$ and $\beta$ 
are large enough, then
there exists a constant $C_{14}>0$, independent of $\beta$, 
$k_n$ and $\Lambda\supset \Lambda(L_n)$, such that
$$
-[f_\Lambda^2]^{(k_n)}_{\,\mu^*}\geq 
C_{14}^{k_n}\,k_n!^\frac{d}{d-1}\,
\Delta^{k_n}\beta^{-\frac{k_n}{d-1}}\,{\chi_2^\prime}^{-\frac{dk_n}{d-1}}\,.
$$

\end{proof}

\subsection{Lower bounds of the derivatives of the free 
energy at infinite volume}\label{subsection2.5}
We show that we can interchange the thermodynamic limit 
and the operation of taking the derivatives, and that
the Taylor series, which exists,  has a radius of convergence equal to $0$.
These statements are a consequence of Lemmas \ref{lem4.2} and \ref{lem4.3}.

\begin{lemma}\label{lem4.2}
If $\beta$ is sufficiently large,
then  for any $k\in\N$ there exists $M_k=M_k(\beta)<\infty$, such that for all 
$t\in(\mu^*-\varepsilon,\mu^*]$ and for all finite $\Lambda$,
\begin{equation}\nonumber
\big|[f^2_{\Lambda}]^{(k)}_{\,t}\big|\leq M_k \,.
\end{equation}
\end{lemma}

\begin{proof}
$\omega(\gd)$ is analytic and $\tau_1(\beta,\theta^\prime)$-stable on a disc of radius $\theta\Delta^{-1}R_2(V(\gd))$. 
From Cauchy formula
$$
\big|[u_\Lambda(\gd)]^{(k)}_{\,t}\big|
\leq k!\,C_{15}^k|\gd|^\frac{k}{d-1}\,{\rm e}^{-\beta\kappa|\gd|}\,,
$$
for some constants $C_{15}$ and $\kappa>0$.
Therefore
\begin{align*}
|\Lambda|\,\beta\,\big|[f^2_{\Lambda}]^{(k)}_{\,t}\big|
&\leq
\sum_{\gd\subset\Lambda}\big|[u_\Lambda(\gd)]^{(k)}_{\,t}\big|
\leq
k!\,C_{15}^k
\sum_{\gd\subset\Lambda}|\gd|^{\frac{k}{d-1}}\,{\rm e}^{-\beta\kappa|\gd|}
\equiv
|\Lambda|\beta M_k\,.
\end{align*}
\end{proof}

\begin{lemma}\label{lem4.3}
$$
\lim_{L\ra\infty}[f^2_{\Lambda(L)}]^{(k)}_{\,\mu^*}=
\lim_{t\uparrow \mu^*}[f]^{(k)}_{\,t}\,.
$$
\end{lemma}

\begin{proof}
We compute the first derivative at the origin. Let $\eta>0$.
\begin{align} 
A(\eta):
&=
\frac{f(\mu^*)-f(\mu^*-\eta)}{\eta}\nonumber \\
&=\lim_{L\ra\infty}
\frac{f^2_{\Lambda(L)}(\mu^*)-f^2_{\Lambda(L)}(\mu^*-\eta)}{\eta} \nonumber\\ 
&=\lim_{L\ra\infty}
\frac{[f^2_{\Lambda(L)}]^{(1)}_{\,\mu^*}\,\eta + 
\frac{1}{2!} [f^2_{\Lambda(L)}]^{(2)}_{\mu^*-x_L(\eta)}\,\eta^2}{\eta}\nonumber\\
&=
\lim_{L\ra\infty}
\Big([f^2_{\Lambda(L)}]^{(1)}_{\,\mu^*}+ 
\frac{1}{2!} [f^2_{\Lambda(L)}]^{(2)}_{\mu^*-x_L(\eta)}\,\eta\Big)\,.
\nonumber
 \end{align}
By Lemma \ref{lem4.2}, $|[f^2_{\Lambda(L)}]^{(2)}_{\mu^*-x_L(\eta)}|\leq M_2$. Therefore $\{A(\eta)\}_\eta$ is a Cauchy sequence. Hence the following limits exist,
$$
[f]^{(1)}_{\,\mu^*}=\lim_{\eta \downarrow 0} \frac{f(\mu^*)-f(\mu^*-\eta)}{\eta}=
\lim_{t\uparrow \mu^*}[f]^{(1)}_{\,t}=
\lim_{L\ra\infty}[f^2_{\Lambda(L)}]^{(1)}_{\,\mu^*}\,.
$$
Same proof for the derivatives of any order.

\end{proof}

\end{document}

%% file: integrale2.pstex_t
\begin{picture}(0,0)%
\includegraphics{integrale2.pstex}%
\end{picture}%
\setlength{\unitlength}{3108sp}%
\begingroup\makeatletter\ifx\SetFigFont\undefined%
\gdef\SetFigFont#1#2#3#4#5{%
  \reset@font\fontsize{#1}{#2pt}%
  \fontfamily{#3}\fontseries{#4}\fontshape{#5}%
  \selectfont}%
\fi\endgroup%
\begin{picture}(3924,4126)(2464,-5039)
\put(4609,-1794){\makebox(0,0)[lb]{\smash{\SetFigFont{9}{10.8}{\rmdefault}{\mddefault}{\updefault}{\color[rgb]{0,0,0}$\partial D_r^d$}%
}}}
\put(2830,-2431){\makebox(0,0)[lb]{\smash{\SetFigFont{9}{10.8}{\rmdefault}{\mddefault}{\updefault}{\color[rgb]{0,0,0}$\partial D_r^g$}%
}}}
\put(4668,-3330){\makebox(0,0)[lb]{\smash{\SetFigFont{9}{10.8}{\rmdefault}{\mddefault}{\updefault}{\color[rgb]{0,0,0}$r$}%
}}}
\put(3862,-1108){\makebox(0,0)[lb]{\smash{\SetFigFont{9}{10.8}{\rmdefault}{\mddefault}{\updefault}{\color[rgb]{0,0,0}$\mu^*(\nu;\beta)$}%
}}}
\put(2729,-1403){\makebox(0,0)[lb]{\smash{\SetFigFont{9}{10.8}{\rmdefault}{\mddefault}{\updefault}{\color[rgb]{0,0,0}$\nu$}%
}}}
\put(6388,-3011){\makebox(0,0)[lb]{\smash{\SetFigFont{9}{10.8}{\rmdefault}{\mddefault}{\updefault}{\color[rgb]{0,0,0}$\mu$}%
}}}
\put(4339,-2881){\makebox(0,0)[lb]{\smash{\SetFigFont{9}{10.8}{\rmdefault}{\mddefault}{\updefault}{\color[rgb]{0,0,0}$\mu^*(0;\beta)$}%
}}}
\put(3234,-4981){\makebox(0,0)[lb]{\smash{\SetFigFont{9}{10.8}{\rmdefault}{\mddefault}{\updefault}{\color[rgb]{0,0,0}$\theta\triangle^{-1}R_1(V(\Gamma^2))$}%
}}}
\put(4087,-4544){\makebox(0,0)[lb]{\smash{\SetFigFont{9}{10.8}{\rmdefault}{\mddefault}{\updefault}{\color[rgb]{0,0,0}$\theta\triangle^{-1}R_2(V(\Gamma^2))$}%
}}}
\end{picture}